\documentstyle[11pt,epsfig]{article}

\newcommand \be{\begin{equation}}
\newcommand \ba{\begin{eqnarray}}
\newcommand \ee{\end{equation}}
\newcommand \ea{\end{eqnarray}}

\begin{document}

\begin{center}
{\LARGE Generalized $q$-Analysis of Log-Periodicity:\\ Applications
to Critical Ruptures}
\end{center}
\bigskip
\begin{center}
{\large Wei-Xing Zhou {\small$^{\mbox{\ref{igpp}}}$} and Didier
Sornette {\small$^{\mbox{\ref{igpp},\ref{ess},\ref{lpec}}}$}}
\end{center}
\bigskip
\begin{enumerate}
\item Institute of Geophysics and Planetary Physics, University of
California, Los Angeles, CA 90095\label{igpp}

\item Department of Earth and Space Sciences, University of
California, Los Angeles, CA 90095\label{ess}

\item Laboratoire de Physique de la Mati\`ere Condens\'ee, CNRS UMR
6622 and Universit\'e de Nice-Sophia Antipolis, 06108 Nice Cedex
2, France\label{lpec}
\end{enumerate}
%%\date{today}

\begin{abstract}

We introduce a generalization of the $q$-analysis, which provides
a novel non-parametric tool for the description and detection of
log-periodic structures associated with discrete scale invariance.
We use this generalized $q$ analysis to construct a signature
called the $(H,q)$-derivative of discrete scale invariance, which
we use to detect the log-periodicity in the energy release rate
and its cumulative preceding the rupture of five pressure tanks
made of composite carbon-matrix material. We investigate the
significance level of the spectral Lomb periodogram of the optimal
$(H,q)$-derivative. We confirm and strengthen previous parametric
results that the energy release rate and it cumulative exhibit
log-periodicity before rupture. However, our tests to use this
method as a scheme for the prediction of the critical value of the
stress at rupture are not encouraging.

\end{abstract}

\section{Introduction}
\label{s:intro}

The fracture of materials is a catastrophic phenomenon of
considerable technological and scientific importance. However, a
reliable identification of precursory signatures of impending
failure is lacking in most cases. A notable exception has been
found \cite{Anifrani,Anifrani2} in the analysis of acoustic
emissions recorded during the pressurization of spherical tanks of
kevlar or carbon fibers pre-impregnated in a resin matrix wrapped
up around a thin metallic liner (steel or titanium) fabricated and
instrumented by A\'erospatiale-Matra Inc. (now EADS). A recent
thorough analysis \cite{Critrup} of the seven acoustic emission
recordings of seven pressure tanks, that was brought to rupture,
has unambiguously characterized the acceleration of acoustic
energy rate $dE/dt$ and found it to be in agreement with a power
law ``divergence'' expected from the critical point theory
proposed in Ref.~\cite{Vanneste}. In addition, strong evidence of
log-periodic corrections was found \cite{Critrup}, that quantify
the intermittent succession of accelerating bursts and quiescent
phases of the acoustic emissions on the approach to rupture.
Ref.~\cite{Critrup} also proposed an improved model accounting for
the cross-over from the non-critical to the critical region close
to the rupture point exhibits interesting predictive potential.
The critical rupture concept, confirmed by other experiments
\cite{Ciliberto}, opens the road towards industrial applications
involving heterogeneous materials such as fiber composites, rocks,
concrete under compression and materials with large distributed
residual stresses \cite{Andersen1}.

However, a time-to-failure behavior following a power law $dE/dt
\propto (t_c-t)^{-\alpha}$ does not provide a reliable and unique
signature: fits of noisy data by such power laws are notoriously
unreliable; for instance, an error of $1\%$ in the determination
of $t_c$ usually leads to errors of tens of percent for the
exponent $\alpha$. In addition, the determination of $t_c$ is very
sensitive to the presence of noise. In order to improve the
determination of $t_c$, the existence of log-periodic oscillations
have been found useful \cite{Anifrani,canonical,Arbati} and have
been used for the implementation of prediction schemes
\cite{Anifrani,Anifrani2,Critrup} with reasonable success.

The demonstration of the existence of log-periodic corrections to
the power law is of both fundamental and practical interest. From
a fundamental point of view, log-periodicity signals a spontaneous
hierarchical organization of damage with an approximate
geometrical set of characteristic scales. A possible mechanism
involves a succession of ultraviolet instabilities of the
Mullins-Sekerka type \cite{HuangpreMS} (see also
\cite{logperiodicreport} for a review). More generally, the
presence of log-periodicity signals the partial breaking of
continuous scale invariance into discrete scale invariance, which
requires that the underlying field theory be non-unitary
\cite{Salsor}. From a practical view point, log-periodicity may
help locking in the fit on experimental data to obtain a better
precision on the recovery of the critical rupture time $t_c$
\cite{Anifrani,Anifrani2,Critrup}.

However, most of the evidence of log-periodicity in rupture
results from parametric fits of the experimental or numerical data
by a log-periodic power law formula, except for
Ref.~\cite{canonical} which introduced a ``canonical'' averaging
method to extract the log-periodic signal directly. Parametric
fits suffer from two problems: (1) the formula cannot avoid some
simplification which for instance omits the presence of harmonics
and/or other structures; (2) parametric fits are delicate due to
possible degeneracies and, in addition, their statistical
significance (i.e., added value) is difficult to estimate. It is
thus important to develop further non-parametric tests. This is
our goal here to present a novel non-parametric method, that turns
out to be very powerful in identifying log-periodicity in noisy
data.

Our method is based on the concept of a $q$-derivative, the
inverse of the Jackson $q$-integral \cite{Jacksonqder}, which is
the natural tool \cite{Erzan1,ErzanEck} for describing  discrete
scale invariance. Indeed, $q$-derivatives can be identified with
the generators of fractal and multifractal sets with discrete
dilatation symmetries \cite{ErzanEck}. Nowhere differentiable
functions which characterize fractal or multifractal sets turn out
to be perfectly well behaved under the $q$-derivative. Discrete
renormalization group equations, whose general mathematical
solutions are power laws with complex exponents (and hence exhibit
log-periodicity), can be seen as nothing but Jackson $q$-integrals
of regular functions of the decimated degrees of freedom. Jackson
$q$-integrals constitutes the natural generalization of regular
integrals for discretely self-similar systems \cite{ErzanEck}. The
way that Jackson $q$-integral can be related to the free energy of
a spin system on a hierarchical lattice was explained in
\cite{Erzan1}.

In section 2, we introduce the $q$-derivative and generalize it to
take into account of anomalous scaling. We discuss the main
properties of the generalized $q$-derivative that will be useful
for our analysis of this paper. Section 3 presents our analysis
with the generalized $q$-derivative of the acoustic emission data,
both for the energy release rate and for the cumulative energy
release, obtained during the pressurization of spherical tanks of
kevlar or carbon fibers pre-impregnated in a resin matrix wrapped
up around a thin metallic liner (steel or titanium). For
comparison, we use exactly the same set of seven acoustic emission
recordings of the seven pressure tanks used in the previous study
reported in \cite{Critrup}. Our new results confirm strongly the
existence of log-periodicity with a much enhanced confidence.
Section 4 tests a scheme using the generalized $q$-derivative for
prediction purpose. Here, we find disappointing results: the
parametric approach in \cite{Critrup} turns out to be more
powerful.

\section{The generalized $q$-analysis and log-periodicity}
\label{s:HqA}

\subsection{Definition}

Let us take some $q\in (0,1)\cup (1,\infty)$.
The $q$-derivative of an arbitrary function $f(x)$ is defined as
\begin{equation}
D_q f(x) = \frac{f(x)-f(qx)}{(1-q)x}~. \label{Eq:qD}
\end{equation}
For $q \to 1$, the definition (\ref{Eq:qD})
recovers the usual definition of a derivative.

For $q \neq 1$, $D_q f(x)$ is more than just a derivative: it
compares the relative variations of $f(x)$ and of $x$ when $x$ is
magnified by the finite factor $q$. It is thus intuitive that the
$q$-derivative tests the scale invariance property of the function
$f(x)$. As we said above, it was actually shown by Erzan and
Eckmann \cite{ErzanEck} that the $q$-derivative is the natural
tool for describing  discrete scale invariance, since a fixed
finite $q$ compares $f(x)$ with $f(qx)$ at $x$ magnified by a
fixed factor, and thus it also compares $f(qx)$ with $f(q^2x)$,
$f(q^2x)$ with $f(q^3x)$, etc. When $x$ is taken as the distance
from a critical point, $D_q f(x)$ thus quantifies the discrete
self-similarity of the function $f(x)$ in the vicinity of the
critical point. From the definition (\ref{Eq:qD}), it is clear
that
\begin{equation}
D_{1/q} f(x) = D_{q} f(x/q)~. \label{Eq:map}
\end{equation}
It is thus enough to study $D_q f(x)$ for $q \in (0,1)$ to derive
its values for all $q$'s.

The necessary and sufficient condition
for a function $f(x)$ of order $\psi$ to be homogeneous is
\begin{equation}
D_q f(x) = \frac{q^\psi-1}{q-1} \frac{f(x)}{x}~.
\label{Eq:HomoCond}
\end{equation}
This expression suggests the introduction of a generalized
$q$-derivative that we call $(H,q)$-derivative, such that the
dependence in $x$ of $D_q^H f(x)$ disappears for homogeneous
functions, for the choice $H=\psi$. Consider therefore the
following definition \be D_q^H f(x) \stackrel{\triangle}{=}
{f(x)-f(qx) \over [(1-q)x]^H}~, \label{Eq:HqD} \ee such that
$D_q^{H=1} f(x)$ recovers the standard $q$-derivative $D_q f(x)$.
For a power law function $f(x) = B x^m$, $D_q^{H=m} [B x^m] = B
(1-q^m)/(1-q)^m$ is constant. For a statistically homogeneous
function $f(x) \stackrel{d}{=} B x^m$, $D_q^{H=1} f(x)
\stackrel{d}{=}$ constant.

The generalized $(H,q)$-derivative has two control parameters: the
discrete scale factor $q$ devised to characterize the log-periodic
structure and the exponent $H$ introduced to account for a
possible power law dependence, i.e., to trends in log-log plots.

\subsection{Application to log-periodic functions}

Since Erzan and Eckmann \cite{ErzanEck} showed that the
$q$-derivative is the natural tool for describing discrete scale
invariance, it is natural to study the properties of the $(H,
q)$-derivative of the simplest function exhibiting discrete scale
invariance, namely a power law decorated by a simple log-periodic
function:
\begin{equation}
y(x) = A - B\tau^mx^m + C\tau^mx^m \cos\left(\omega\ln{x} \right)~,
\label{Eq:Sor}
\end{equation}
where the presence of a phase $\phi = \omega \ln{\tau}$ has been
absorbed in the definition of $x$, $0 < C < B$ and $\omega=2\pi
f$. This equation, where $x$ is interpreted as the normalized
distance $x= (p_c-p)/\tau$ to a critical point $p_c$, has been
used in several works to describe material rupture
\cite{Anifrani,Anifrani2,Arbati,Critrup}, precursory patterns of
large earthquakes \cite{Sorsam}, rock bursts \cite{ouillonrock},
aftershocks \cite{Afterhuang,LeeSornette}, and speculative bubbles
preceding financial crashes \cite{financecrash}.

The $(H,q)$-derivative of $y(x)$ is
\begin{equation}
D_q^H y(x) = x^{m-H} [B' + C'g(x)]~, \label{Eq:DqHlogcos}
\end{equation}
where
\be
B' =-{ B\tau^m(1-q^m) \over (1-q)^H}, ~~~~~~~~C'={C\tau^m \over (1-q)^H}
\label{mgjjbvd}
\ee
  and
  \be g(x) = C_1 \cos(\omega \ln{x}) +
C_2\sin(\omega\ln{x})~, \label{Eq:bracket} \ee with $C_1=1 -
q^m\cos(\omega \ln{q})>0$ and $C_2=\sin(\omega\ln{q})$. The
special choice $H=m$ gets rid of the power law and the
$(H,q)$-derivative $D_q^H y(x)$ becomes the pure log-periodic
function $g(x)$.

For the special choices $q=e^{-n 2 \pi/\omega}$ where $n$ is a
positive integer corresponding to choosing $q$ equal to one of the
preferred scaling factor of the log-periodic function, we obtain
$C_1=1 - q^m$ and $C_2=0$. Therefore, the $(H, q)$-derivative
offers a novel approach for detecting the preferred scaling
factors of the discretely scale invariant function by a measure of
its phase: those values of $q$ such that the phases of the $(H=m,
q)$-derivative is that of a pure cosine should qualify as the
preferred scaling factors. Such phases can be for instance
measured by the Hilbert transform. Here, we do not persue this
possibility which wll be explored in another presentation.

In the Appendix A, it is shown that $g(x)$ is extremal at $x_m$
solution of
\begin{equation}
\omega \ln(x_m) = n\pi + \arctan(C_2/C_1)~,
\end{equation}
where $n = 0,\pm 1,\pm 2,\cdots$, and that the extreme
values of $g(x)$ are
\begin{equation}
g(x_m) = \pm \sqrt{C_1^2+C_2^2}~.
\end{equation}
The amplitude of $D_q^H y(x)$ is then
\begin{equation}
A = x_m^{m-H}~C'~\sqrt{C_1^2+C_2^2}~, \label{Eq:Amp1}
\end{equation}
while the successive extreme values are
\begin{equation}
D_m = \left(B' \pm C'\sqrt{C_1^2+C_2^2}\right)(x_m)^{m-H}~,
\label{Eq:Dm1}
\end{equation}
where $B'$ and $C'$ are defined in (\ref{mgjjbvd}).
By fixing $H$ close to $m$, one can in principle obtain the
amplitude $A_m$ to be constant as a function of $x$. This value
$H=m$ should provide theoretically the most significant
log-periodic component, quantified for instance by the largest
peak of the Lomb periodogram. However, in practice, the noise
embedded in the data may distort the log-periodic oscillations and
the most significant log-periodicity may occur for $H \neq m$. The
introduction of $H$ affords a convenient detrending scheme.

\begin{figure}
\begin{center}
\epsfig{file=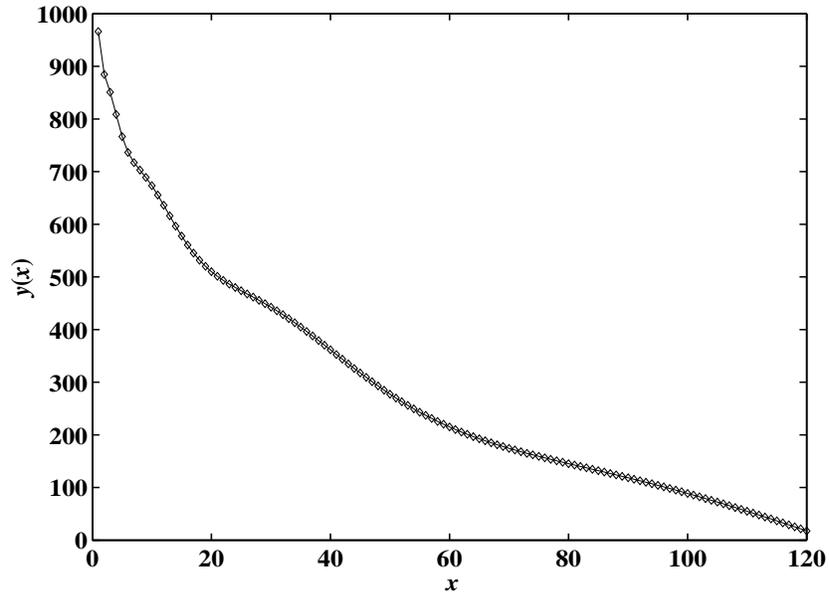,width=11cm, height=8cm}
\end{center}
\caption{Plot of $y(x)$ defined by Eq.~(\ref{Eq:Sor}) as a
function of the pressure-to-rupture $x$ with $A = 1260$, $B=300$,
$C=6$, $m = 0.3$, $\omega = 5.4$ and $\phi=0$. We generated 120
evenly spaced data points with $x$ between $1$ and $120$, to mimic
the cumulative energy release of a real acoustic emission
experiment. } \label{Fig:SorFun}
\end{figure}

\begin{figure}
\begin{center}
\epsfig{file=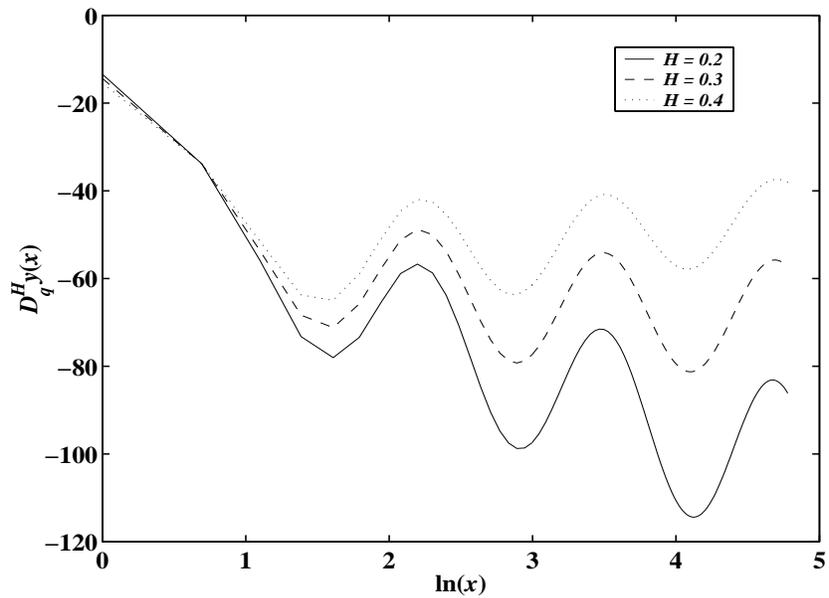,width=11cm, height=8cm}
\end{center}
\caption{The generalized $q$-derivative of $y(x)$ shown in
Fig.~\ref{Fig:SorFun} with $q=0.5$ for $H=0.2$, $H=0.3$ and
$H=0.4$, respectively. The calculation assumes a critical point at $x_c=1$
while the synthetic function has its genuine critical point at $x_c=0$.
This incorrect value of the critical point in the calculation of the
generalized
$q$-derivative is responsible for the distortions observed for $x<1$.
} \label{Fig:SorHqDy}
\end{figure}

Figure \ref{Fig:SorFun} shows $y(x)$ defined in Eq.~(\ref{Eq:Sor})
as a function of the distance $x$ to the critical point with $A =
1260$, $B=300$, $C=6$, $m = 0.3$, $\omega = 5.4$ and $\phi=0$.
Figure~\ref{Fig:SorHqDy} shows its generalized $q$-derivative with
$q=0.5$ for $H=0.2$, $H=0.3$ and $H=0.4$, respectively. This generalized
$q$-derivative has been calculated by using the incorrect assumption that
the critical point is at $x=1$ in order to also show the distortion
resulting from an error in the determination of the critical point.
This distortion becomes important when $x$ is not large compared to $1$.
We observe that the amplitude of the oscillations of $D_q^H y(x)$
increases with $H$ when going towards the critical point $x=0$,
in agreement with the prediction (\ref{Eq:Amp1}).

In the next section, using the $(H,q)$-analysis, we test for the
presence of log-periodicity in the energy release rate and its
cumulative obtained in the experimental recordings of the seven
pressure tanks used in the previous study reported in
\cite{Critrup}.

\section{Investigation of the significance of
log-periodicity in acoustic emission energy release using the
generalized $(H,q)$-analysis} \label{s:revisit}

We use the same notations to label the data sets as used in
Ref.~\cite{Critrup}. In this analysis, we use the true value $p_c$
of the pressure at which rupture occurred in order to define the
pressure-to-failure $p_c-p$ quantifying the distance to the
critical rupture point. We follow \cite{Critrup} and use the
acoustic emission data from the pressure at which a noticeable
acceleration in the cumulative energy release takes place. This
choice is not crucial at all, since the $(H,q)$-analysis is not
sensitive to the points far from the critical point. We also
exclude the last six points nearest to $p_c$ because they contain
the largest noise and may suffer from finite size effects that may
lead to serious distortions. In our analysis, we exclude data sets
of pressure tanks $\#\,5$ and $\#\,7$, because for these two data
sets,  $p_{\mathtt{last}} \ll p_c$ (the last data point is for a
pressure far below the critical rupture value). This leads to few
oscillations and low statistical significance. We thus are left to
analyze five data sets $\#\,1$, $\#\,2$, $\#\,3$, $\#\,4$ and
$\#\,6$.

\subsection{Energy release rate}
\label{ss:form}

In each analysis, $q$ ranges from $0.1$ to $0.9$ with spacing
$0.1$, while $H$ takes values from $-0.9$ to $0.9$ with spacing
$0.1$. This defines $9 \times 19 = 171$ parameter pairs $(H,q)$.
For each parameter pair $(H,q)$, we calculate the generalized
$q$-derivative of the energy release rate. We then perform a
spectral analysis of the generalized $q$-derivative using the Lomb
periodogram method \cite{Numrec}, in order to test for the
statistical significance of possible log-periodic oscillations.
For each $(H,q)$ pair, the highest peak $P_N(H,q)$ and its
associated logarithmic angular frequency $\omega(H,q)$ in the Lomb
periodogram are obtained. The {\bf first criterion} used to
identify a log-periodic signal is the strength of the Lomb
periodogram analysis, i.e., the height of the spectral peaks.

\begin{figure}
\begin{center}
\epsfig{file=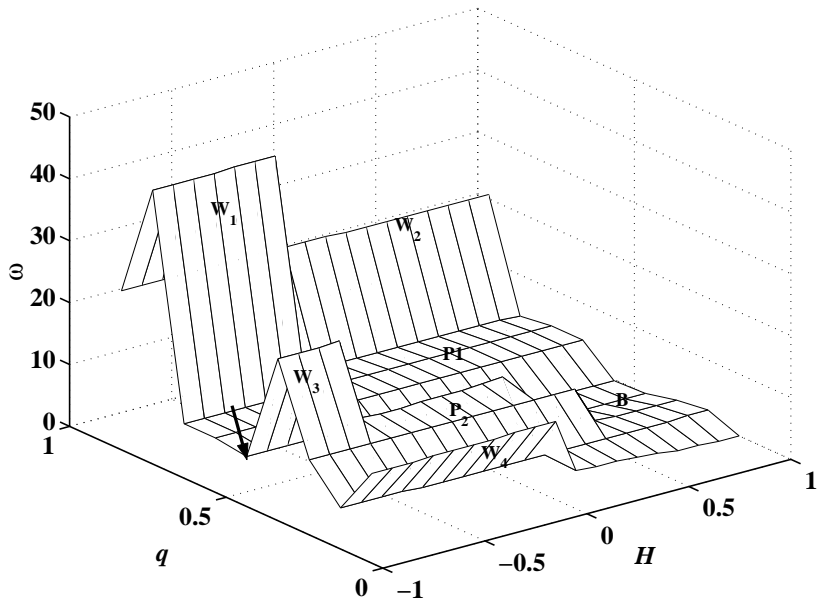,width=11cm, height=8cm}
\end{center}
\caption{Dependence of the logarithmic angular frequency
$\omega(H,q)$ of the most significant peak in each Lomb
periodogram of the $(H,q)$-derivative of the energy release rate
before the rupture of tank $\#\,1$. The regions ${\mathbf{W_1}}$,
${\mathbf{W_2}}$, ${\mathbf{W_3}}$, ${\mathbf{W_4}}$ and
${\mathbf{B}}$ are excluded and the optimal pair $(-0.9,0.5)$ is
indicated by an arrow in the platform ${\mathbf{P_1}}$.}
\label{Fig:DSIForm01f}
\end{figure}

\begin{figure}
\begin{center}
\epsfig{file=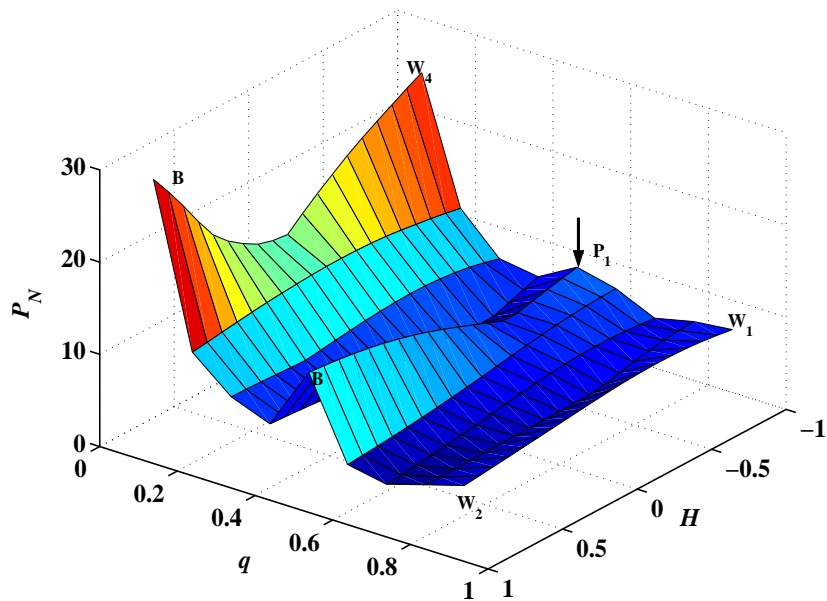,width=11cm, height=8cm}
\end{center}
\caption{Dependence of the height $P_N(H,q)$ of the most
significant peak in each Lomb periodogram of the
$(H,q)$-derivative of the energy release rate before the rupture
of tank $\#\,1$. The optimal pair $(-0.9,0.5)$ is indicated by an
arrow in the platform ${\mathbf{P_1}}$.} \label{Fig:DSIForm01PN}
\end{figure}

\begin{figure}
\begin{center}
\epsfig{file=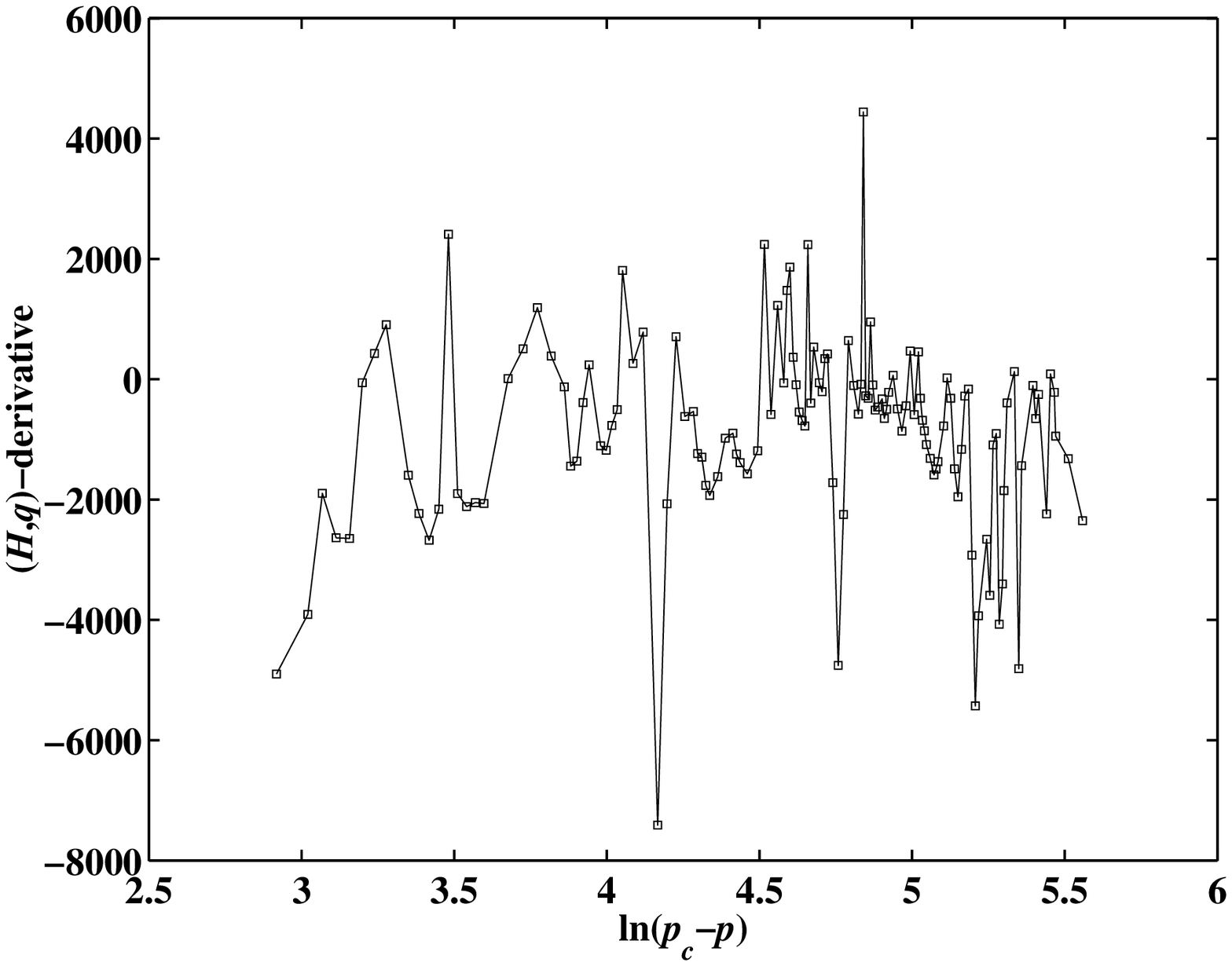,width=12cm, height=9cm}
\end{center}
\caption{$(H,q)$-derivative of the energy release rate before
the rupture of tank $\#\,1$ as a function of the
pressure-to-rupture $p_c-p$ with $q=0.5$ and $H=-0.9$.}
\label{Fig:DSIForm01Dy}
\end{figure}

\begin{figure}
\begin{center}
\epsfig{file=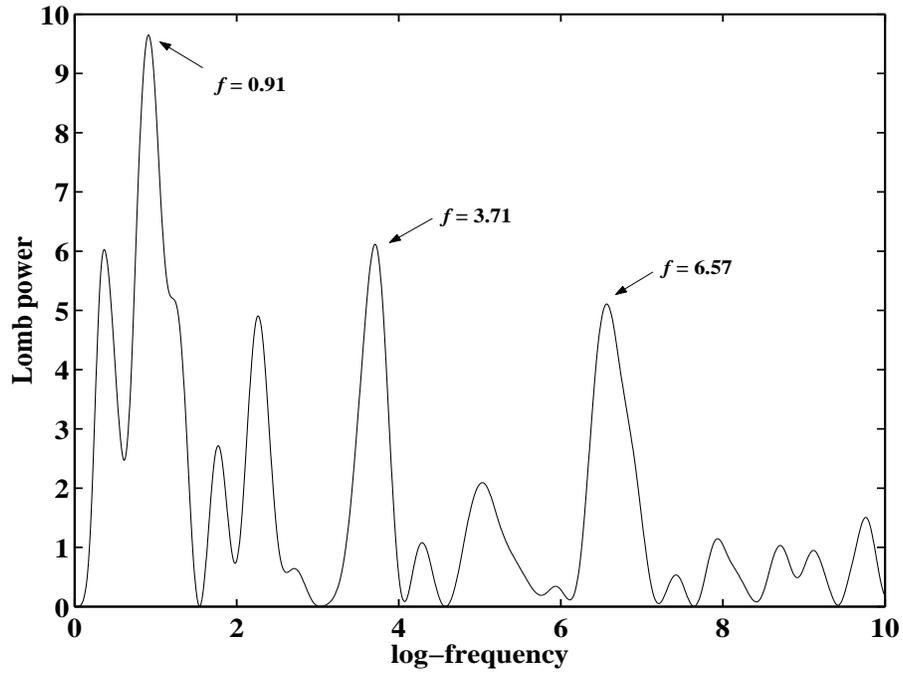,width=12cm, height=9cm}
\end{center}
\caption{Lomb power of the $(H,q)$-derivative shown in
Fig.~\ref{Fig:DSIForm01Dy}. The log-frequency is $0.91$ which has
two clear harmonics at $3.71$ and $6.57$.}
\label{Fig:DSIForm01Lomb}
\end{figure}

\begin{figure}
\begin{center}
\epsfig{file=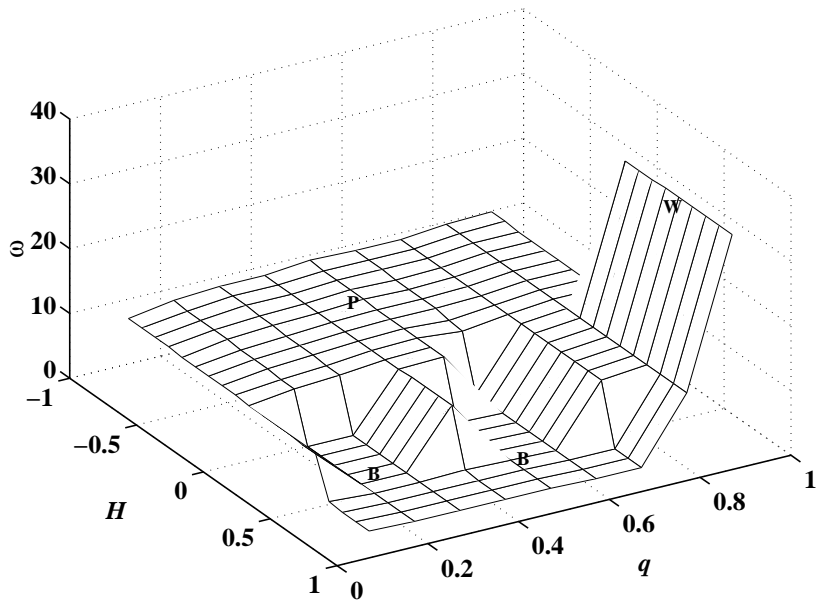,width=11cm, height=8cm}
\end{center}
\caption{Dependence of the logarithmic angular frequency
$\omega(H,q)$ of the most significant peak in each Lomb
periodogram of the $(H,q)$-derivative of the energy release rate
before the rupture of tank $\#\,2$. The wall ${\mathbf{W}}$ is
excluded by both the third and the fourth criteria, while
${\mathbf{B}}$ is excluded by the second criterion.}
\label{Fig:DSIForm02f}
\end{figure}

\begin{figure}
\begin{center}
\epsfig{file=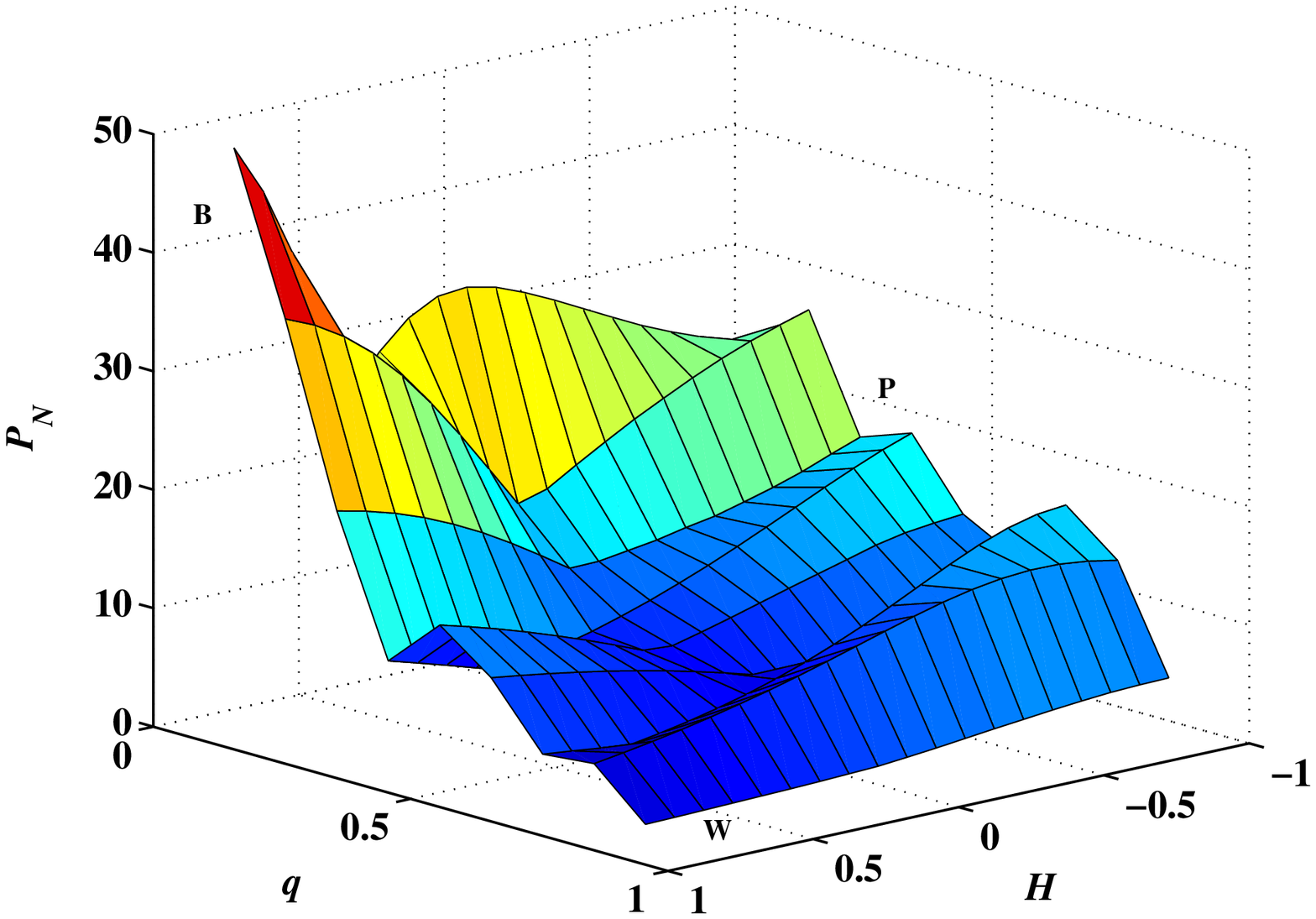,width=11cm, height=8cm}
\end{center}
\caption{Dependence of the height $P_N(H,q)$ of the most
significant peak in each Lomb periodogram of the
$(H,q)$-derivative of the energy release rate before the rupture
of tank $\#\,2$.} \label{Fig:DSIForm02PN}
\end{figure}

\begin{figure}
\begin{center}
\epsfig{file=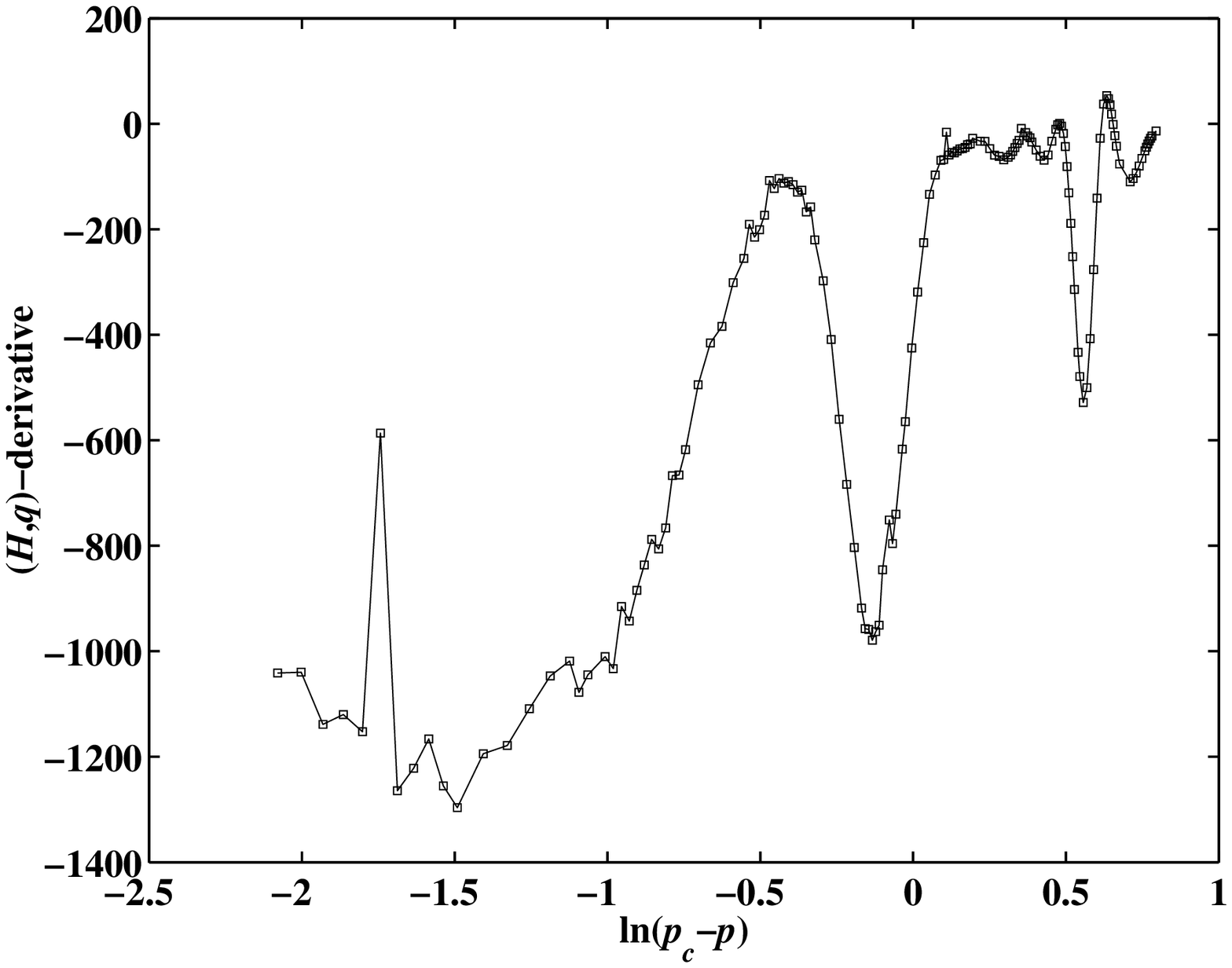,width=12cm, height=9cm}
\end{center}
\caption{$(H,q)$-derivative of the energy release rate before
the rupture of tank $\#\,2$ as a function of the
pressure-to-rupture $p_c-p$ with $q=0.1$ and $H=0.1$.}
\label{Fig:DSIForm02Dy}
\end{figure}

\begin{figure}
\begin{center}
\epsfig{file=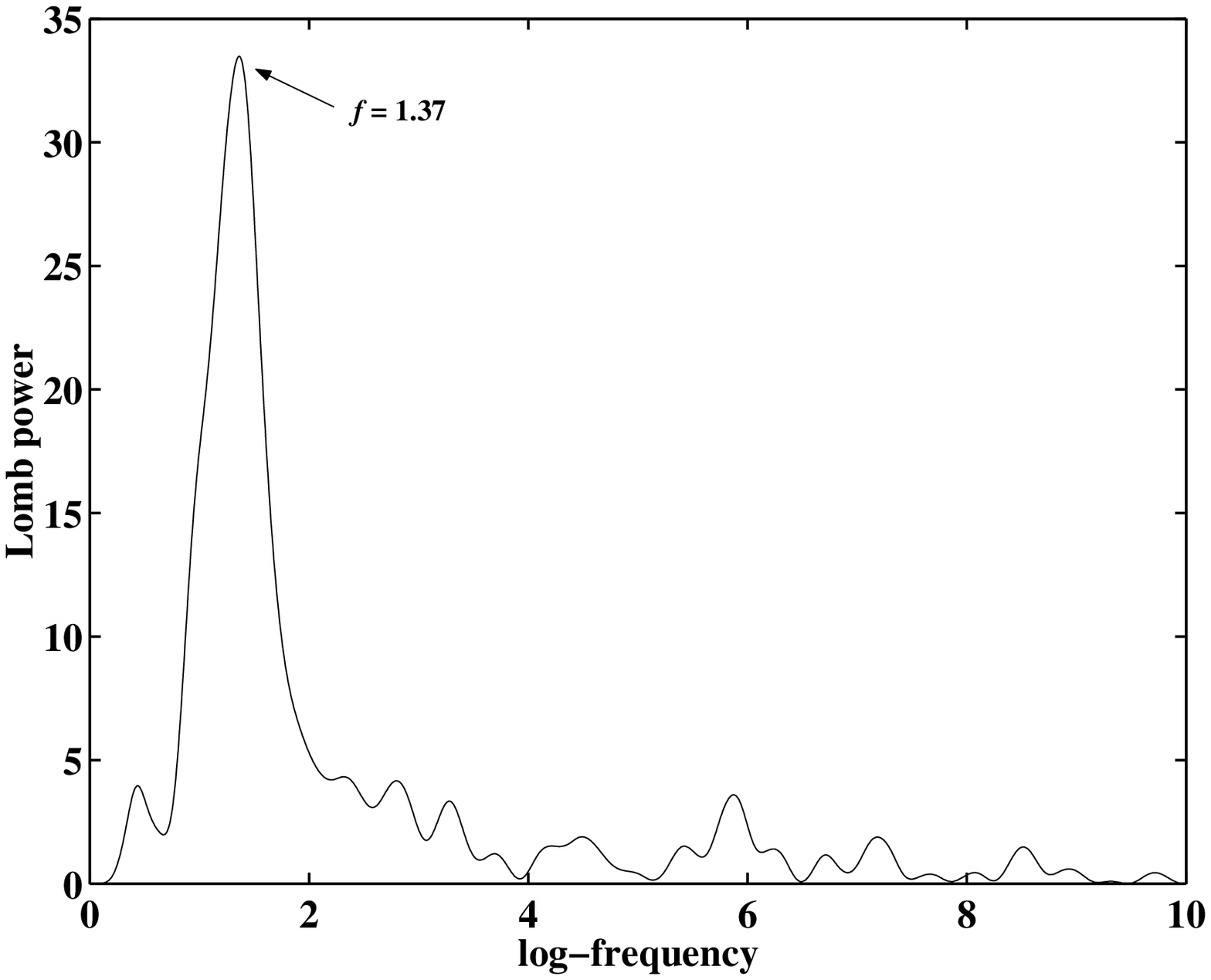,width=12cm, height=9cm}
\end{center}
\caption{Lomb power of the $(H,q)$-derivative shown in
Fig.~\ref{Fig:DSIForm02Dy}.} \label{Fig:DSIForm02Lomb}
\end{figure}

\begin{figure}
\begin{center}
\epsfig{file=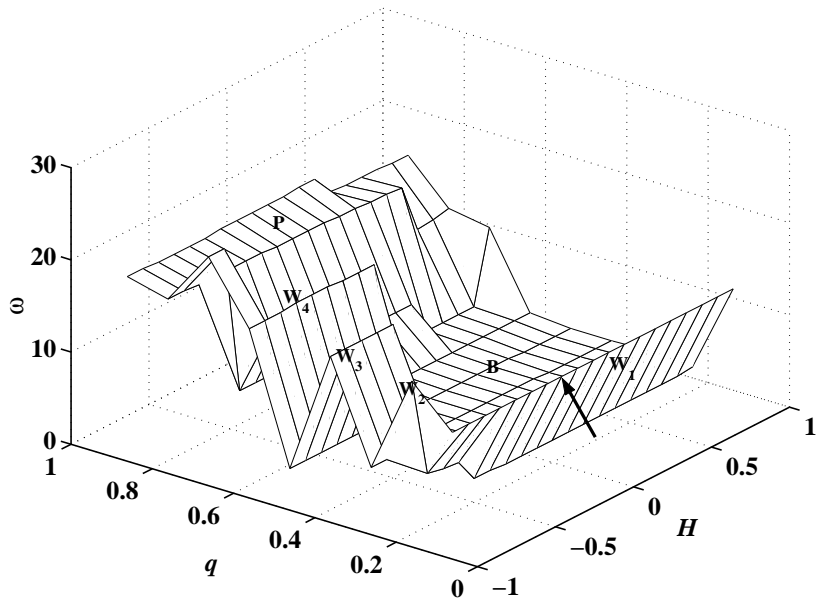,width=11cm, height=8cm}
\end{center}
\caption{Dependence of the logarithmic angular frequency
$\omega(H,q)$ of the most significant peak in each Lomb
periodogram of the $(H,q)$-derivative of the energy release rate
before the rupture of tank $\#\,3$. Regions ${\mathbf{B}}$ and
${\mathbf{W_2}}$, ${\mathbf{W_3}}$, ${\mathbf{W_4}}$ and
${\mathbf{P}}$ are excluded by the second and the third criteria,
respectively. The fourth criterion is not used in this case and
the optimal pair $(-0.2,0.1)$ is at a wall ${\mathbf{W_1}}$ as
indicated by an arrow.} \label{Fig:DSIForm03f}
\end{figure}

\begin{figure}
\begin{center}
\epsfig{file=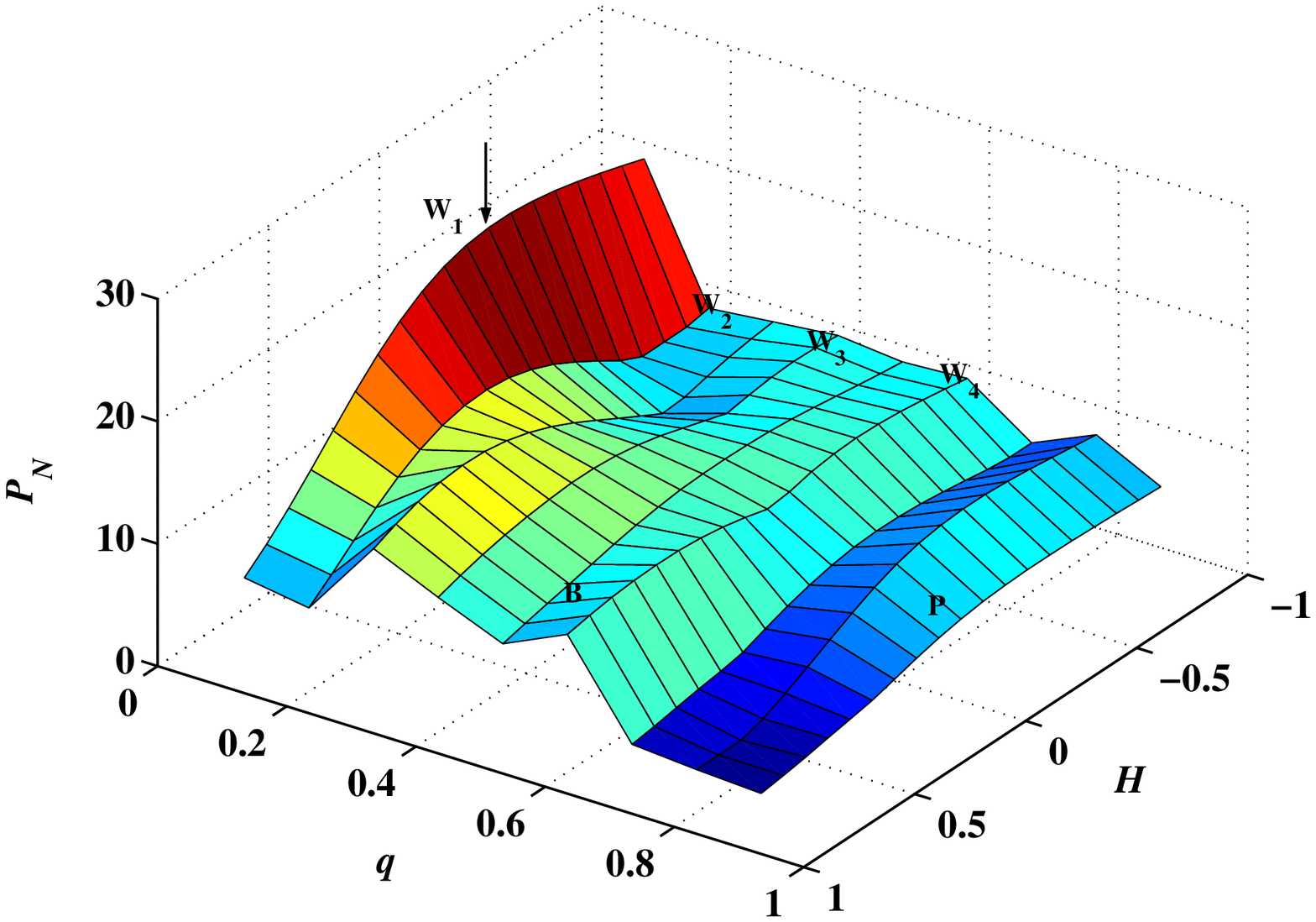,width=11cm, height=8cm}
\end{center}
\caption{Dependence of the height $P_N(H,q)$ of the most
significant peak in each Lomb periodogram of the
$(H,q)$-derivative of the energy release rate before the rupture
of tank $\#\,3$.} \label{Fig:DSIForm03PN}
\end{figure}

\begin{figure}
\begin{center}
\epsfig{file=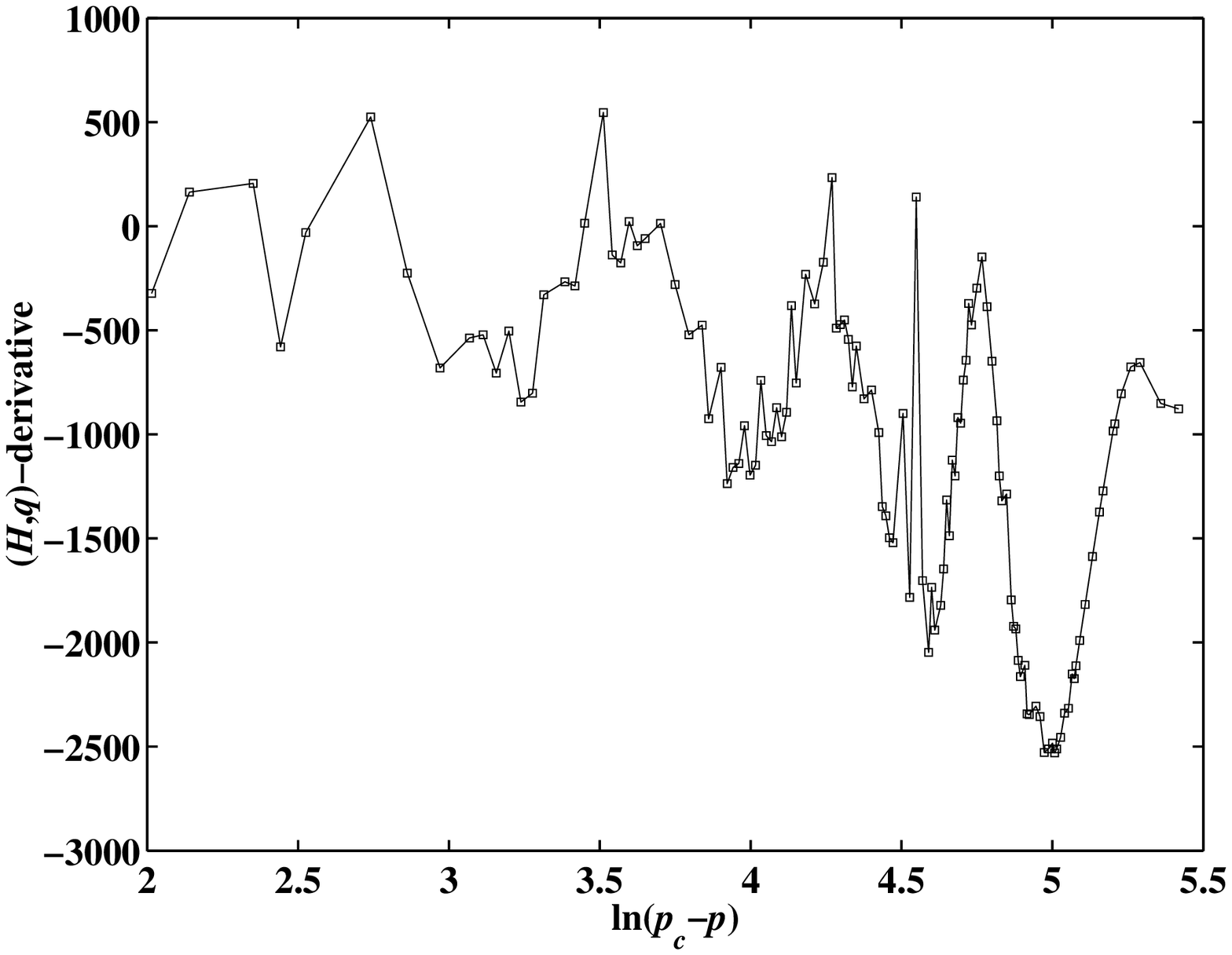,width=12cm, height=9cm}
\end{center}
\caption{$(H,q)$-derivative of the energy release rate before
the rupture of tank $\#\,3$ as a function of the
pressure-to-rupture $p_c-p$ with $q=0.1$ and $H=-0.2$.}
\label{Fig:DSIForm03Dy}
\end{figure}

\begin{figure}
\begin{center}
\epsfig{file=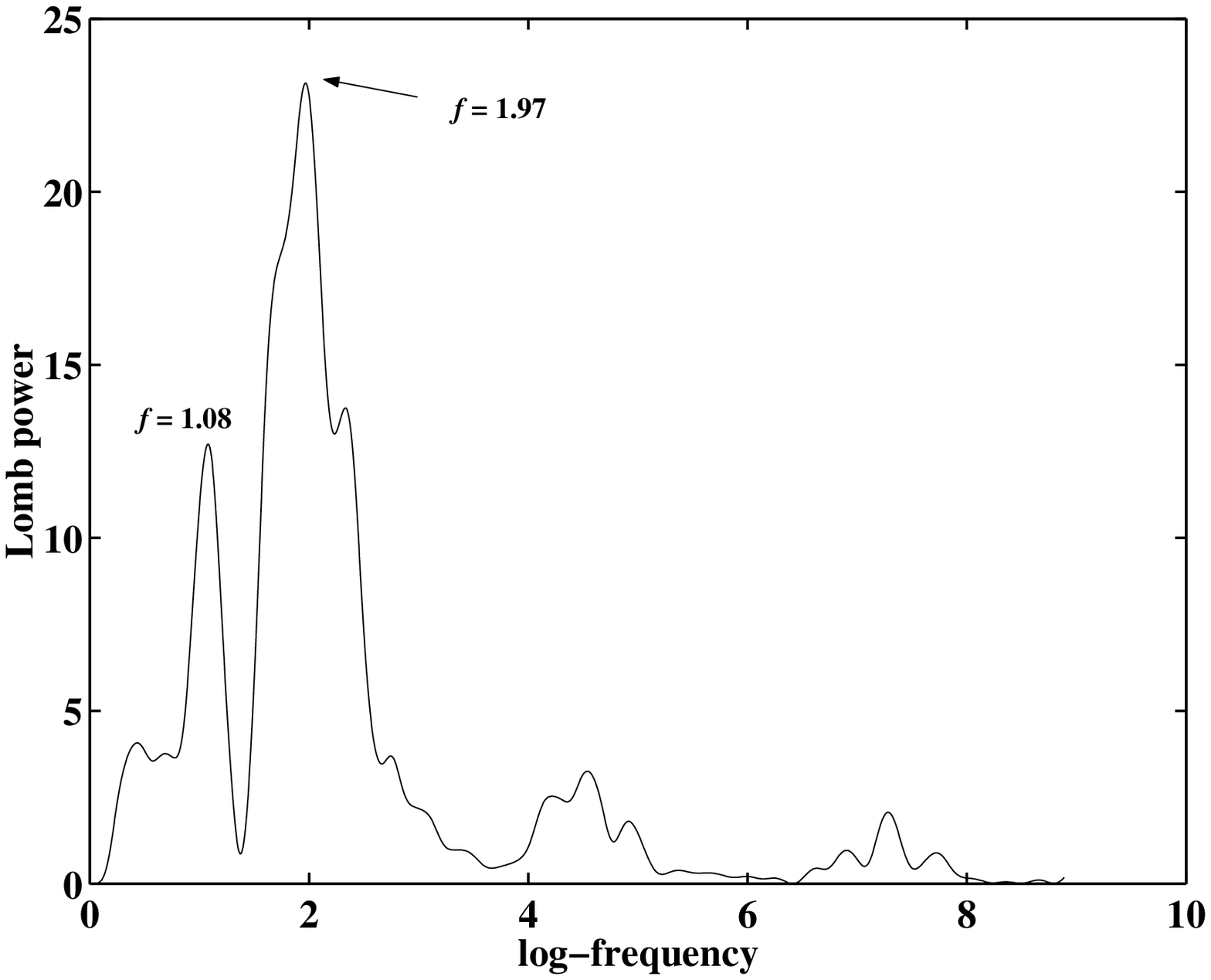,width=12cm, height=9cm}
\end{center}
\caption{Lomb power of the $(H,q)$-derivative shown in
Fig.~\ref{Fig:DSIForm03Dy}.} \label{Fig:DSIForm03Lomb}
\end{figure}

\begin{figure}
\begin{center}
\epsfig{file=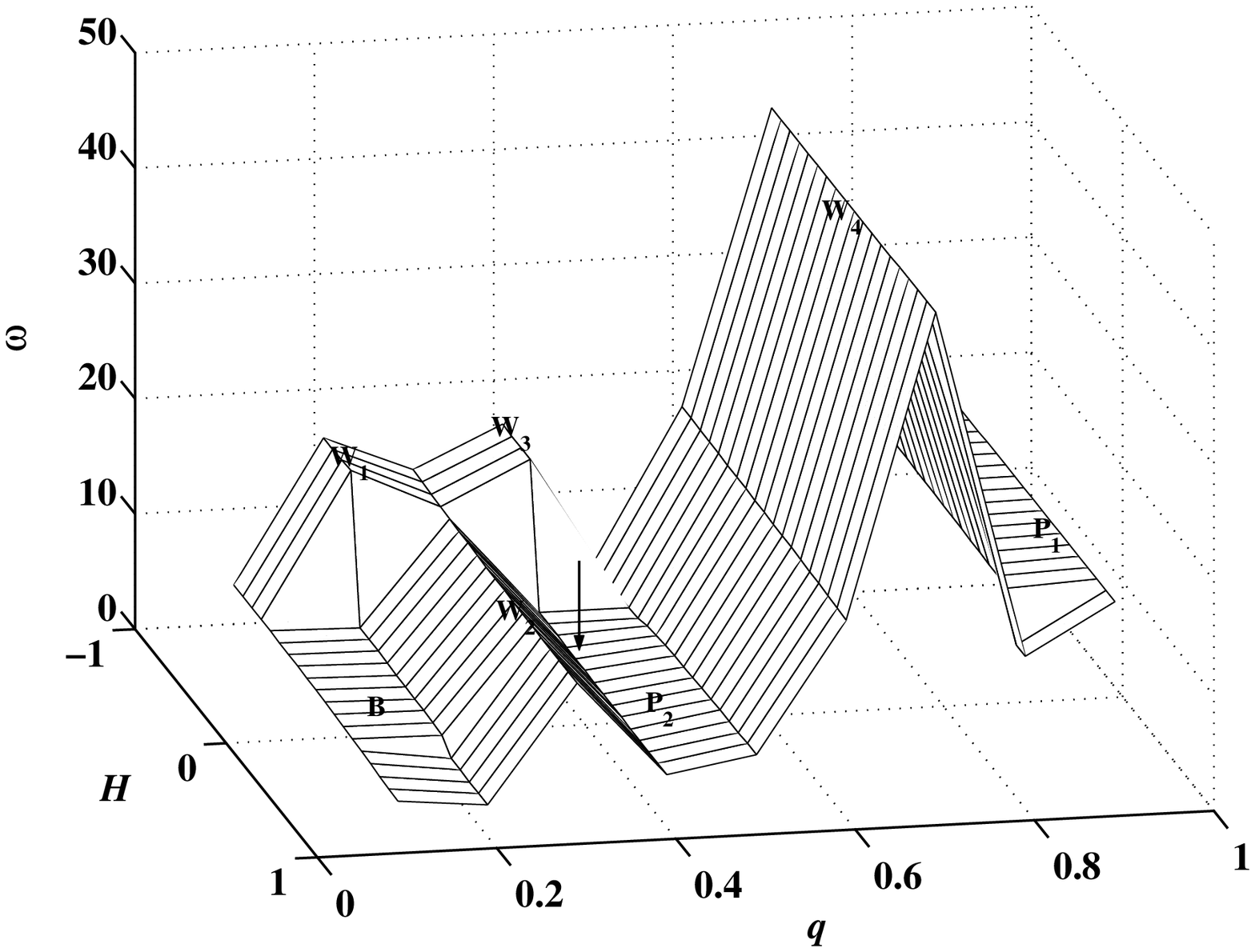,width=11cm, height=8cm}
\end{center}
\caption{Dependence of the logarithmic angular frequency
$\omega(H,q)$ of the most significant peak in each Lomb
periodogram of the $(H,q)$-derivative of the energy release rate
before the rupture of tank $\#\,4$. The optimal pair
$(0.3,0.4)$ is in the platform ${\mathbf{P_2}}$ indicated by an
arrow.} \label{Fig:DSIForm04f}
\end{figure}

\begin{figure}
\begin{center}
\epsfig{file=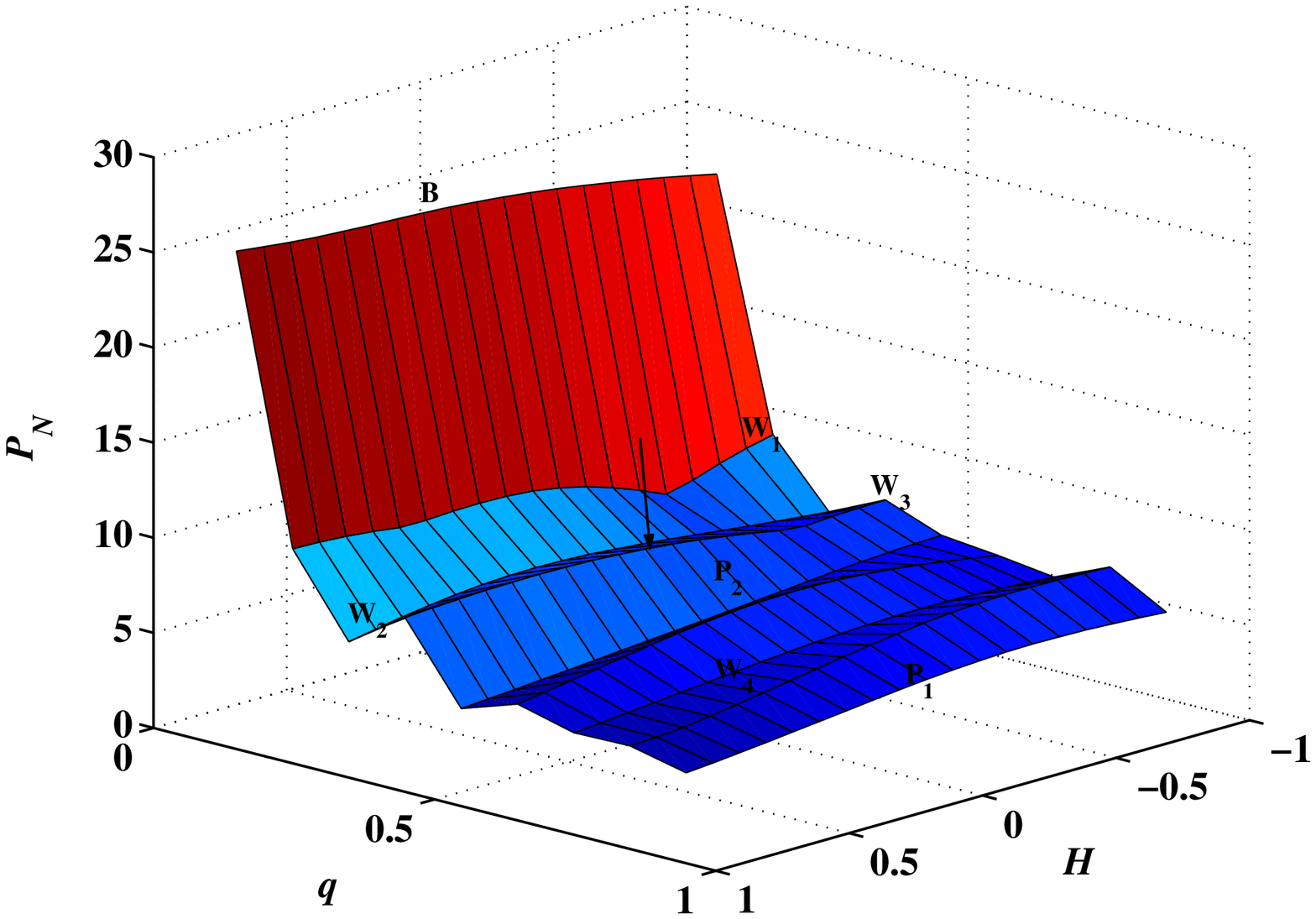,width=11cm, height=8cm}
\end{center}
\caption{Dependence of the height $P_N(H,q)$ of the most
significant peak in each Lomb periodogram of the
$(H,q)$-derivative of the energy release rate before the rupture
of tank $\#\,4$.} \label{Fig:DSIForm04PN}
\end{figure}

\begin{figure}
\begin{center}
\epsfig{file=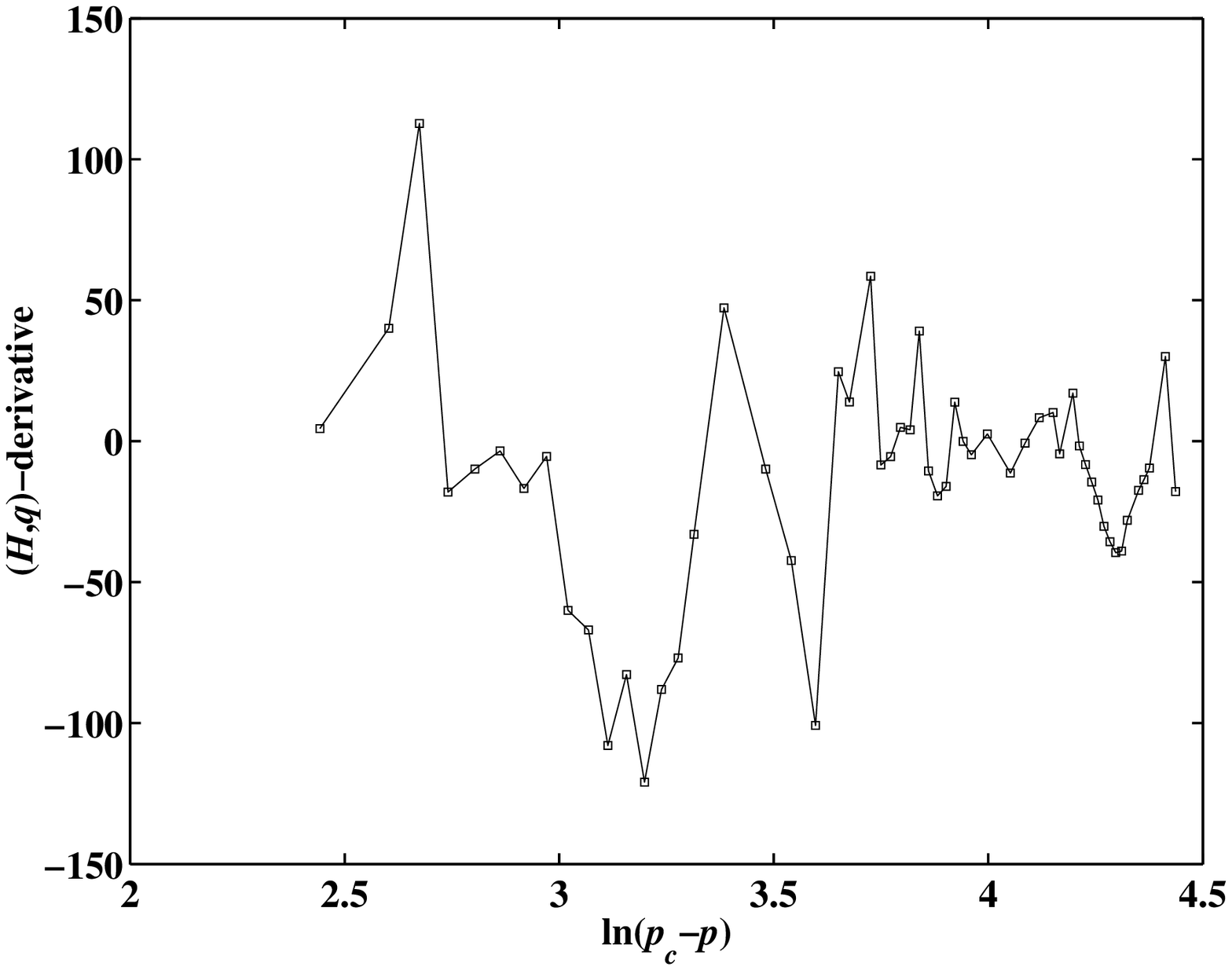,width=12cm, height=9cm}
\end{center}
\caption{$(H,q)$-derivative of the energy release rate before
the rupture of tank $\#\,4$ as a function of the
pressure-to-rupture $p_c-p$ with $q=0.4$ and $H=0.3$.}
\label{Fig:DSIForm04Dy}
\end{figure}

\begin{figure}
\begin{center}
\epsfig{file=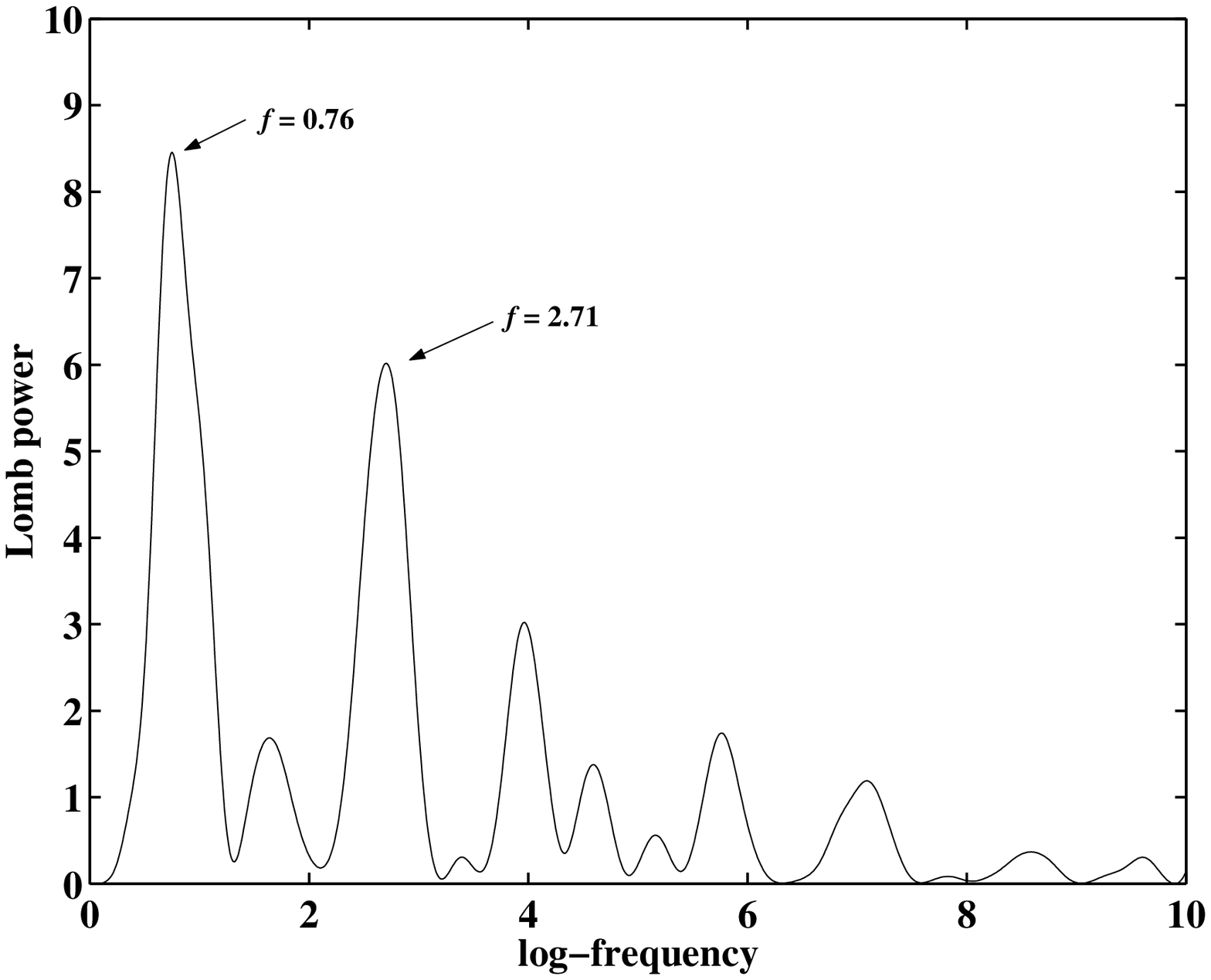,width=12cm, height=9cm}
\end{center}
\caption{Lomb power of the $(H,q)$-derivative shown in
Fig.~\ref{Fig:DSIForm04Dy}. The fundamental log-frequency $f =
0.76$ and its harmonic $f=2.71$ are indicated by arrows.}
\label{Fig:DSIForm04Lomb}
\end{figure}

\begin{figure}
\begin{center}
\epsfig{file=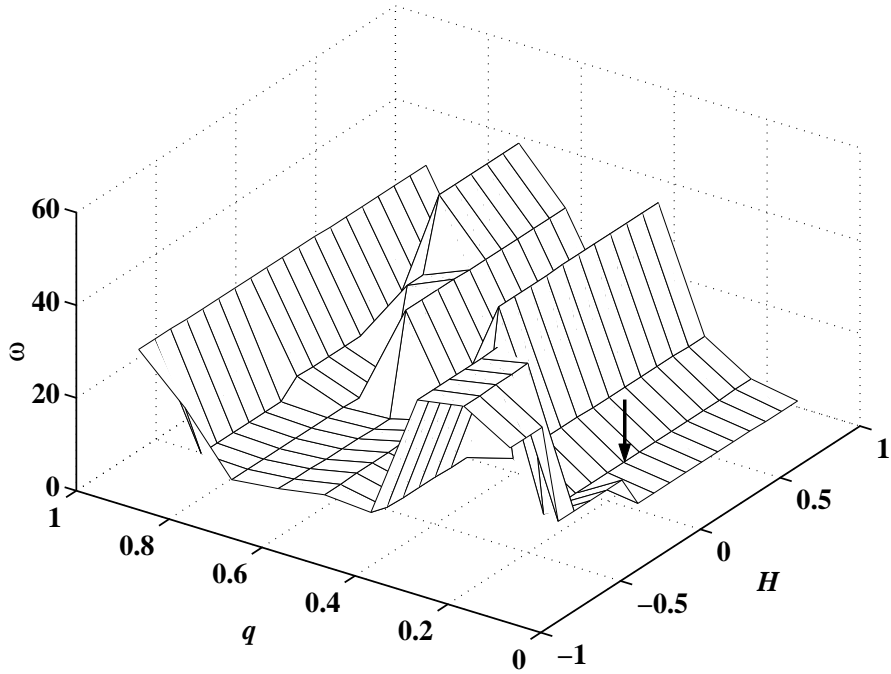,width=12cm, height=9cm}
\end{center}
\caption{Dependence of the logarithmic angular frequency
$\omega(H,q)$ of the most significant peak in each Lomb
periodogram of the $(H,q)$-derivative of the energy release rate
before the rupture of tank $\#\,6$. The optimal pair $(0.1,0.2)$
is indicated by an arrow.} \label{Fig:DSIForm06f}
\end{figure}

\begin{figure}
\begin{center}
\epsfig{file=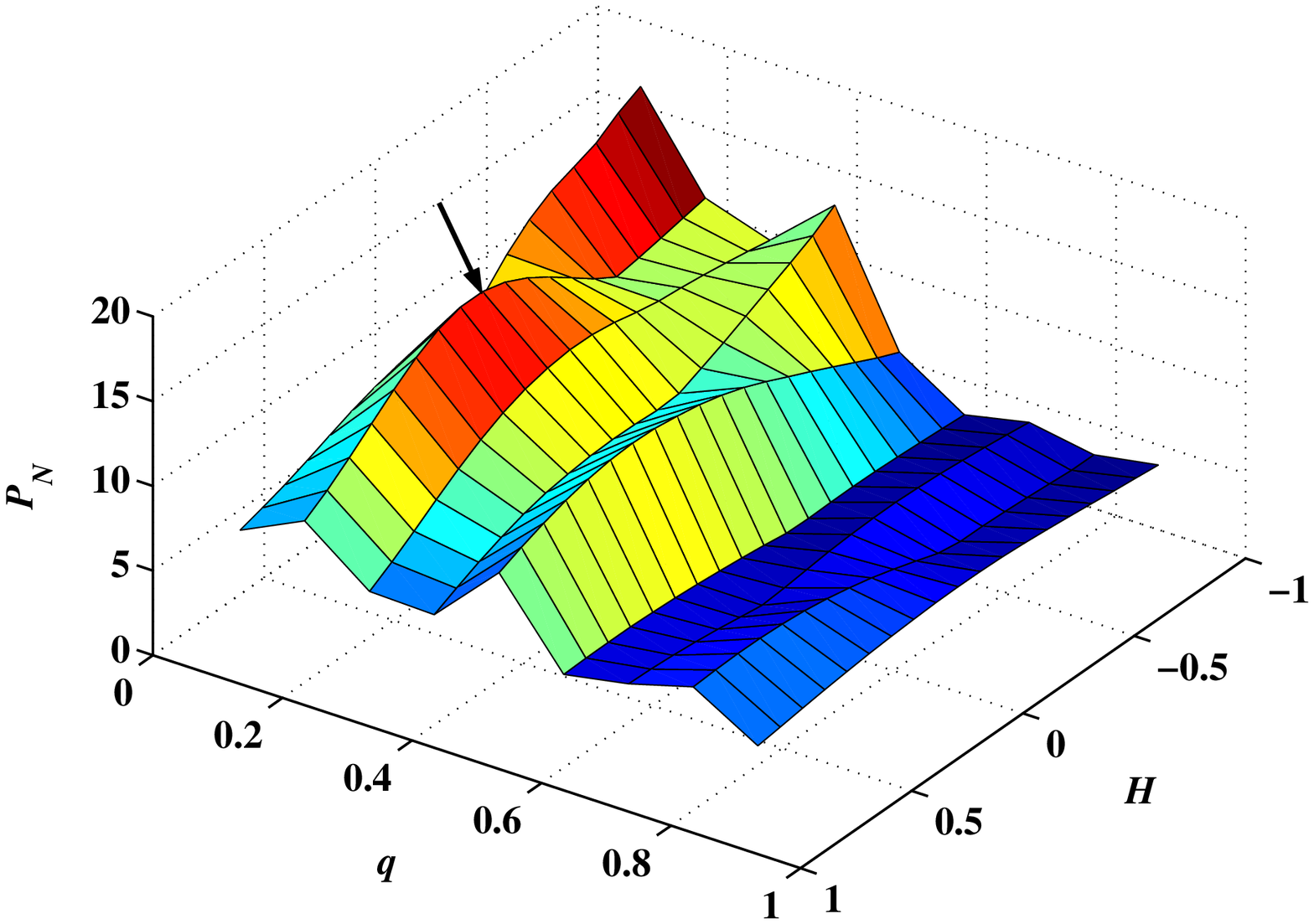,width=12cm, height=9cm}
\end{center}
\caption{Dependence of the height $P_N(H,q)$ of the most
significant peak in each Lomb periodogram of the
$(H,q)$-derivative of the energy release rate before the rupture
of tank $\#\,6$.} \label{Fig:DSIForm06PN}
\end{figure}

\begin{figure}
\begin{center}
\epsfig{file=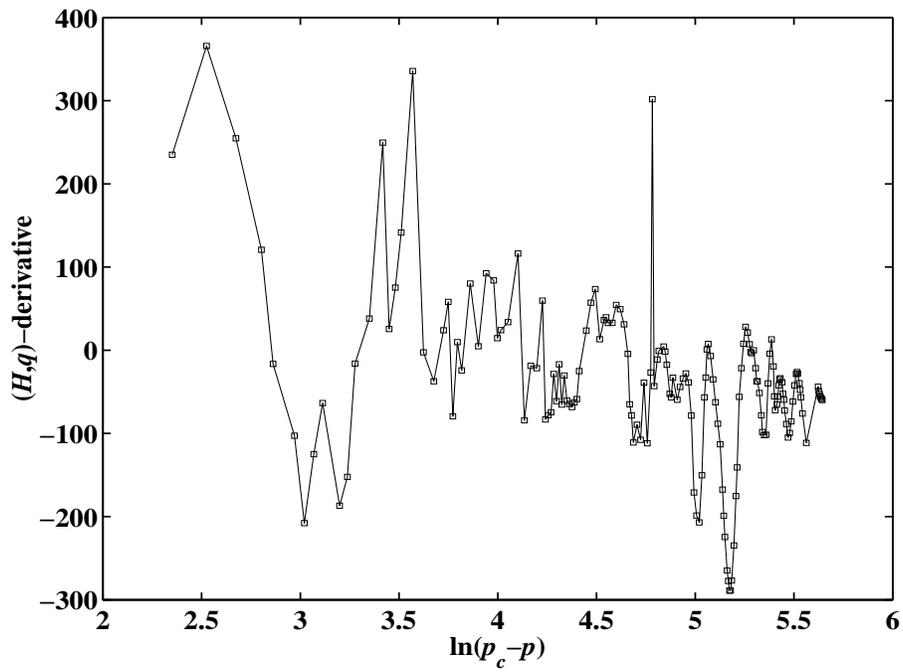,width=12cm, height=9cm}
\end{center}
\caption{$(H,q)$-derivative of the energy release rate before
the rupture of tank $\#\,6$ with respect to the
pressure-to-rupture $p_c-p$ with $q=0.2$ and $H=0.1$.}
\label{Fig:DSIForm06Dy}
\end{figure}

\begin{figure}
\begin{center}
\epsfig{file=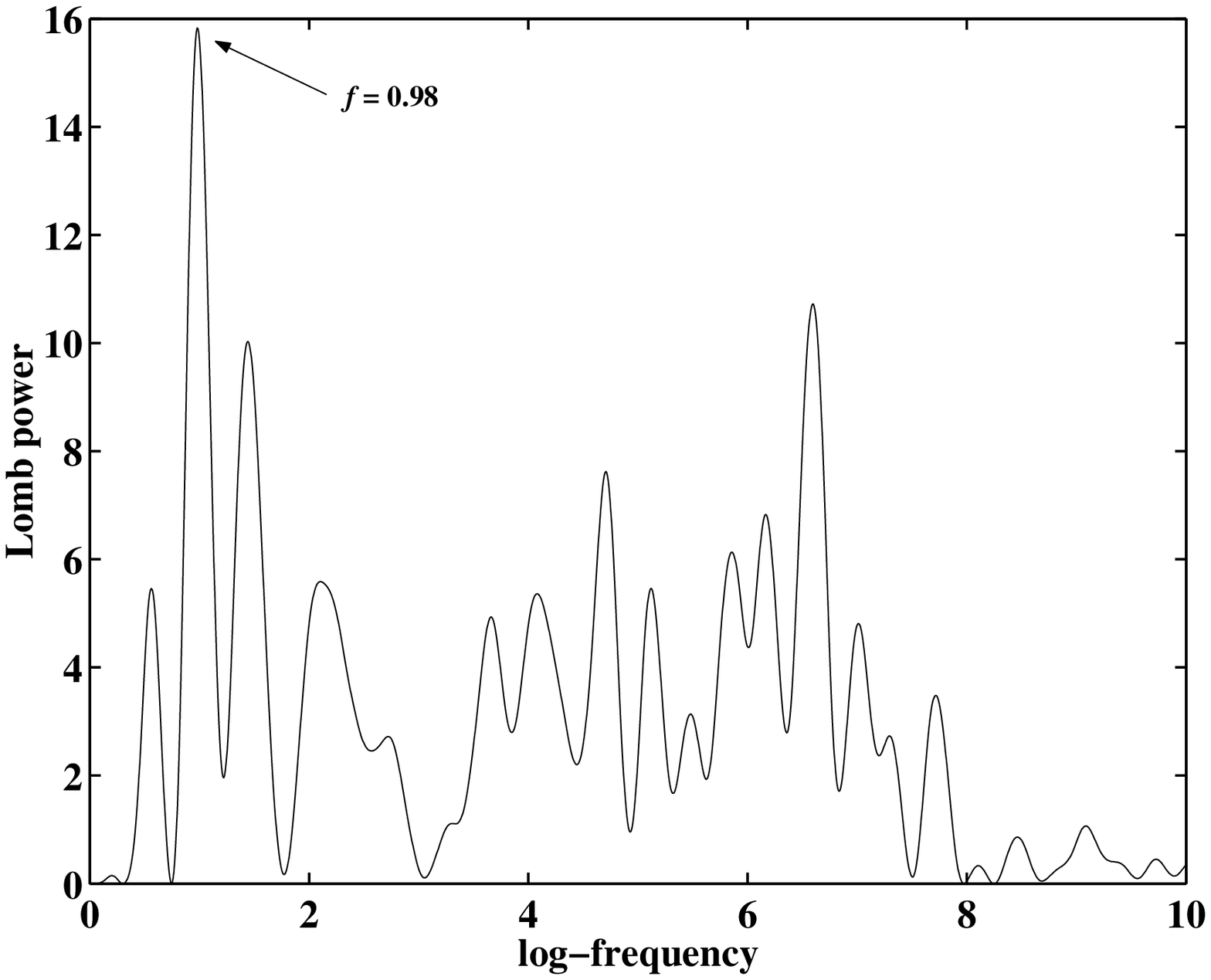,width=12cm, height=9cm}
\end{center}
\caption{Lomb power of the $(H,q)$-derivative shown in
Fig.~\ref{Fig:DSIForm06Dy}.} \label{Fig:DSIForm06Lomb}
\end{figure}

The results for the five data sets $\#\,1$, $\#\,2$, $\#\,3$,
$\#\,4$ and $\#\,6$ are shown in Figs.~\ref{Fig:DSIForm01f} to
\ref{Fig:DSIForm06Lomb}. Figures \ref{Fig:DSIForm01PN},
\ref{Fig:DSIForm02PN}, \ref{Fig:DSIForm03PN},
\ref{Fig:DSIForm04PN} and \ref{Fig:DSIForm06PN} show the
dependence of the highest peak $P_N(H,q)$ in each Lomb periodogram
as a function of $H$ and $q$. Figs.~\ref{Fig:DSIForm01f},
\ref{Fig:DSIForm02f}, \ref{Fig:DSIForm03f}, \ref{Fig:DSIForm04f}
and \ref{Fig:DSIForm06f} give the associated logarithmic angular
frequencies $\omega(H,q)$. The other figures show the generalized
$q$-derivative for the pair $(H,q)$ giving the highest Lomb peak
and the associated Lomb periodogram.

In order to interpret the results presented in the figures
\ref{Fig:DSIForm01f} to \ref{Fig:DSIForm06Lomb}, we need to
explain the criteria that we have used. First, to make the
description simpler and more geometrical, as shown in
Fig.~\ref{Fig:DSIForm01f}, we first classify all pairs of $(H,q)$
into three classes: ${\mathbf{W}}$ (wedge or wall), ${\mathbf{P}}$
(platform) and ${\mathbf{B}}$ (bottom of valley or basin).

In Fig.~\ref{Fig:DSIForm01PN}, the pairs $(H,q)$ near $(0.1,0.1)$
gives the largest $P_N$, which implies the most significant
log-periodic oscillations in the generalized $q$ derivative.
However, the associated $\omega$ is about $3\sim 4$ as shown in
the region ${\mathbf{B}}$ of Fig.~\ref{Fig:DSIForm01f}. The
problem is that this value is dangerously close to the most
probable (log-periodic) angular-frequency $\omega^{mp} = 3.6$,
resulting solely from the most probable noise. It was indeed shown
\cite{Afterhuang} that noise decorating power laws may lead to
artifactual log-periodicity with a most probable frequency
corresponding roughly to $1.5$ oscillations over the whole range
of analysis. In the present context, this defines the most
probable (log-periodic) angular frequency $\omega^{mp}$ by the
following formula \be \omega^{mp} = {2\pi \times 1.5 \over
\ln(p_c-p_{\min})-\ln(p_c-p_{\max})}~, \label{Eq:wmp} \ee where
the acoustic emission signal is recorded from the pressure
$p_{\min}$ to $p_{\max}$. In the case of experiment $\#\,1$, this
leads to $\omega^{mp} = 3.6$. Thus, a value of $\omega$ in the
range $3\sim 4$ corresponds to approximately 1.5 oscillations in
the plot of the generalized $q$-derivative as a function of
$\ln(p_c-p)$ for the whole range of pressure. This is not
sufficient. We should thus exclude these pairs of $(H,q)$. This
defines our {\textbf{second criterion}} for qualifying the
existence of log-periodicity: pairs of $(H,q)$ with $\omega(H,q)
\le \omega^{mp}$ should be discarded, because there is a
non-negligible probability that the observed log-periodicity may
result from noisy fluctuations around the power law and is thus
spurious. Ideally, we should estimate the probability that random
noise, of several plausible standard distributions, creates a
false alarm that a periodicity (or log-periodicity) is found in
the $(H,q)$-derivative of the signal. This has been done in a
systematic way for a large variety of noise, without and with
long-range correlation \cite{zhousorstats}. However, it is
difficult to identify what should be the correct null hypothesis
of the noise decorating the generalized $q$-derivative. For the
sake of simplicity, we thus stick to the simple rule called
{\textbf{second criterion}} just discussed.

A log-periodic angular frequency which is too small may result
from noise, as we just said. Similarly, a too high log-periodic
angular frequency is also the signature of noise, simply because
it is easy to fit noisy data with many oscillations. We should
thus also discard the pairs of $(H,q)$ whose $\omega$ are too
large. Ref.~\cite{Critrup} used this criterion to discard
solutions with $\omega \geq 14$. This gives us a {\textbf{third
criterion}}. It is not always obvious to fix the value of the
threshold angular frequency beyond which solutions are rejected as
noise. Fortunately, this third criterion can often be obtained to
follow from the first criterion, since the Lomb peak $P_N$
corresponding to a large angular frequency is usually low
\cite{zhousorstats}. We apply this criterion and hence exclude
$(H,q)$ pairs in the regions of ${\mathbf{W_1}}$, ${\mathbf{W_2}}$
and ${\mathbf{W_3}}$ shown in Fig.~\ref{Fig:DSIForm01f}.

A physically meaningful optimal pair of $(H,q)$ is thus obtained.
For data set $\#\,1$, it corresponds to $(H,q)=(-0.9,0.5)$, which
is indicated by arrows in Fig.~\ref{Fig:DSIForm01f} and
Fig.~\ref{Fig:DSIForm01PN}. It lies within a platform shown in
Fig.~\ref{Fig:DSIForm01f}. We also show in
Fig.~\ref{Fig:DSIForm01Dy} the generalized $q$-derivative and in
Fig.~\ref{Fig:DSIForm01Lomb} its corresponding Lomb periodogram.
The log-periodic oscillations are found to be very significant.
Under the null hypothesis of independent Gaussian noise decorating
the signal, the false-alarm probability of the log-periodic
component is $\approx 0.06\%$. The false-alarm probability for the
null hypothesis of independent heavy-tailed noises (say, L\'evy
stable noise and power-law noise) and weakly correlated noises
(say, Garch(1,1) noise and fractional Gaussian noise (fGn) with
the Hurst index less than $0.5$)  is even lower
\cite{zhousorstats}. If we assume a strongly correlated noise such
as fractional Gaussian noise with the Hurst index greater than
$0.5$, the false-alarm probability increases: for instance, a
fractional Gaussian noise with Hurst index of $0.62$ leads to a
false-alarm probability of $1\%$ \cite{zhousorstats}.

As a {\textbf{fourth criterion}}, if
the majority of the pairs $(H,q)$ have similar oscillatory
behavior (i.e., similar $\omega$), we take that this indicates
a robust log-periodic signal and that the optimal pair should be
within these pairs with high probability. This excludes
${\mathbf{W_1}}$, ${\mathbf{W_2}}$, ${\mathbf{W_3}}$ and
${\mathbf{W_4}}$ in Fig.~\ref{Fig:DSIForm01f} and strengthens the
choice of the previous criteria. This fourth criterion is sometimes
ambiguous to apply in practice. However, all experimental systems have their
optimal pair  $(H,q)$ in well-defined platforms ${\mathbf{P}}$, except for
experiment $\#\,3$ which presents its optimal $(H,q)$ in ${\mathbf{W_1}}$.

Table~\ref{Tb:form} summurizes the results of the
$(H,q)$-analysis on the energy release rate of data sets $\#\,1$,
$\#\,2$, $\#\,3$, $\#\,4$ and $\#\,6$. For each data set, the
generalized $q$-analysis was performed between $(p_{\min},
p_{\max})$, with a number of points varying between 55 and 164
as indicated in the column ``Points.'' $p_c$ is the true critical
pressure at rupture. $P_N$ is the optimal value of the Lomb peak
height, and $\omega$ is the corresponding logarithmic angular
frequency. The column ``Gaussian'' presents the false-alarm
probability of the Lomb peak under the null hypothesis of independent
Gaussian noise. The column ``fGn'' evaluates the value of the Hurst
index of a fractional Brownian noise
which would give a false-alarm probability $1\%$ to get the same peak
as in our analysis.

These results give a very strong confidence that there is a genuine
log-periodicity of the energy release rate before rupture in the
data sets $\#\,1$, $\#\,2$, $\#\,3$, $\#\,4$ and $\#\,6$. This confirms
and strengthens Ref.~\cite{Critrup}, which claimed that
a pure power law fails to parameterize the
data but a log-periodic formulae does a good job.

\begin{table}
\begin{center}
\begin{tabular}{|c|c|c|c|c|c|c|c|c|c|c|c|c|}
\hline Tank&Points&$p_{\min}$&$p_{\max}$&$p_c$&$(H,q)$&$P_N$
&$\omega$&$\omega '$&$\omega^{mp}$&$N_{osc}$&Gaussian&fGn\\\hline
$\#\,1$&131&453.5&694.5&713&$(-0.9,0.5)$&9.7&5.7&23.3&3.6&2.4&8E-3&0.56\\\hline
$\#\,2$&147&451.5&660.5&673&$(0.1,0.1)$&33.8&8.6&$/$&3.3&3.9&3E-13&0.92\\\hline
$\#\,3$&121&538.5&756.5&764&$(-0.2,0.1)$&23.1&12.4&6.8&2.8&6.6&1E-8&0.
82\\\hline
$\#\,4$&55&671.5&744.5&756&$(0,0.4)$&8.7&4.7&17.0&4.7&1.5&9E-3&0.52\\\hline
$\#\,6$&164&451.5&723.5&734&$(0.1,0.2)$&15.8&6.2&$/$&2.9&3.2&2E-5&0.71\\\hline
\end{tabular}
\caption{Summary of the results of the $(H,q)$-analysis on the
energy release rate of data sets $\#\,1$, $\#\,2$, $\#\,3$,
$\#\,4$ and $\#\,6$. For each data set, the generalized
$q$-analysis was performed between $(p_{\min}, p_{\max})$ with a
number of points given in the column ``Points.'' $p_c$ is the true
critical pressure of rupture. The column $(H,q)$ lists the optimal
pairs. $P_N$ is the optimal value of the Lomb peak height.
$\omega$ and $\omega '$ are the corresponding logarithmic angular
frequency and the angular frequency corresponding to the second
highest peak. $N_{osc}$ is the number of oscillations. The column
``Gaussian'' presents the false-alarm probability of the Lomb peak
under null hypothesis of independent Gaussian noise. The column
``fGn'' gives the Hurst index of a fractional Gaussian noise that
would give a false-alarm probability $1\%$ to obtain the same Lomb
peak as in the signal.} \label{Tb:form}
\end{center}
\end{table}

\subsection{Cumulative energy release}
\label{ss:cumu}

\begin{figure}
\begin{center}
\epsfig{file=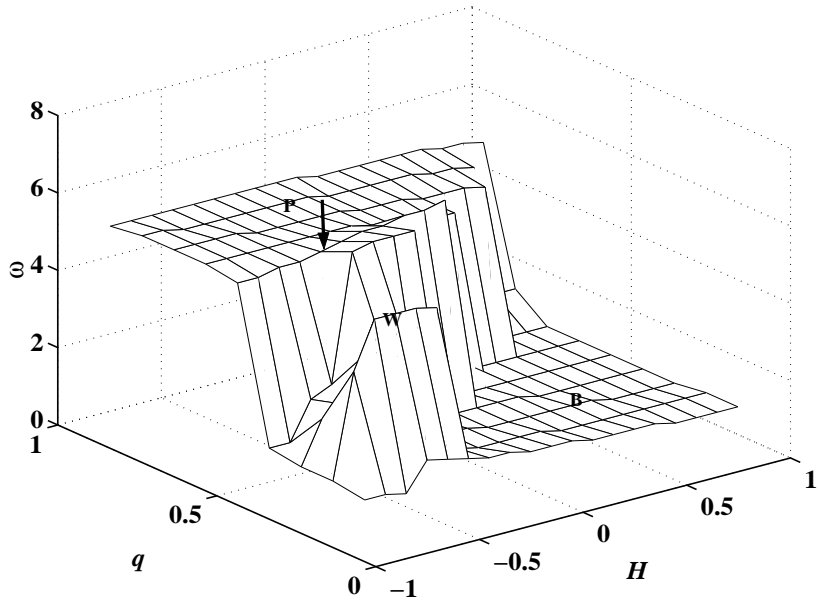,width=11cm, height=8cm}
\end{center}
\caption{Dependence of the logarithmic angular frequency
$\omega(H,q)$ of the most significant peak in each Lomb
periodogram of the $(H,q)$-derivative of the cumulative energy
release before the rupture of tank $\#\,1$. The wedge
${\mathbf{W}}$ and the bottom ${\mathbf{B}}$ are excluded by the
second and fourth criteria. The optimal pair $(-0.5,0.6)$ is
indicated by an arrow in the platform ${\mathbf{P}}$.}
\label{Fig:DSICumu01f}
\end{figure}

\begin{figure}
\begin{center}
\epsfig{file=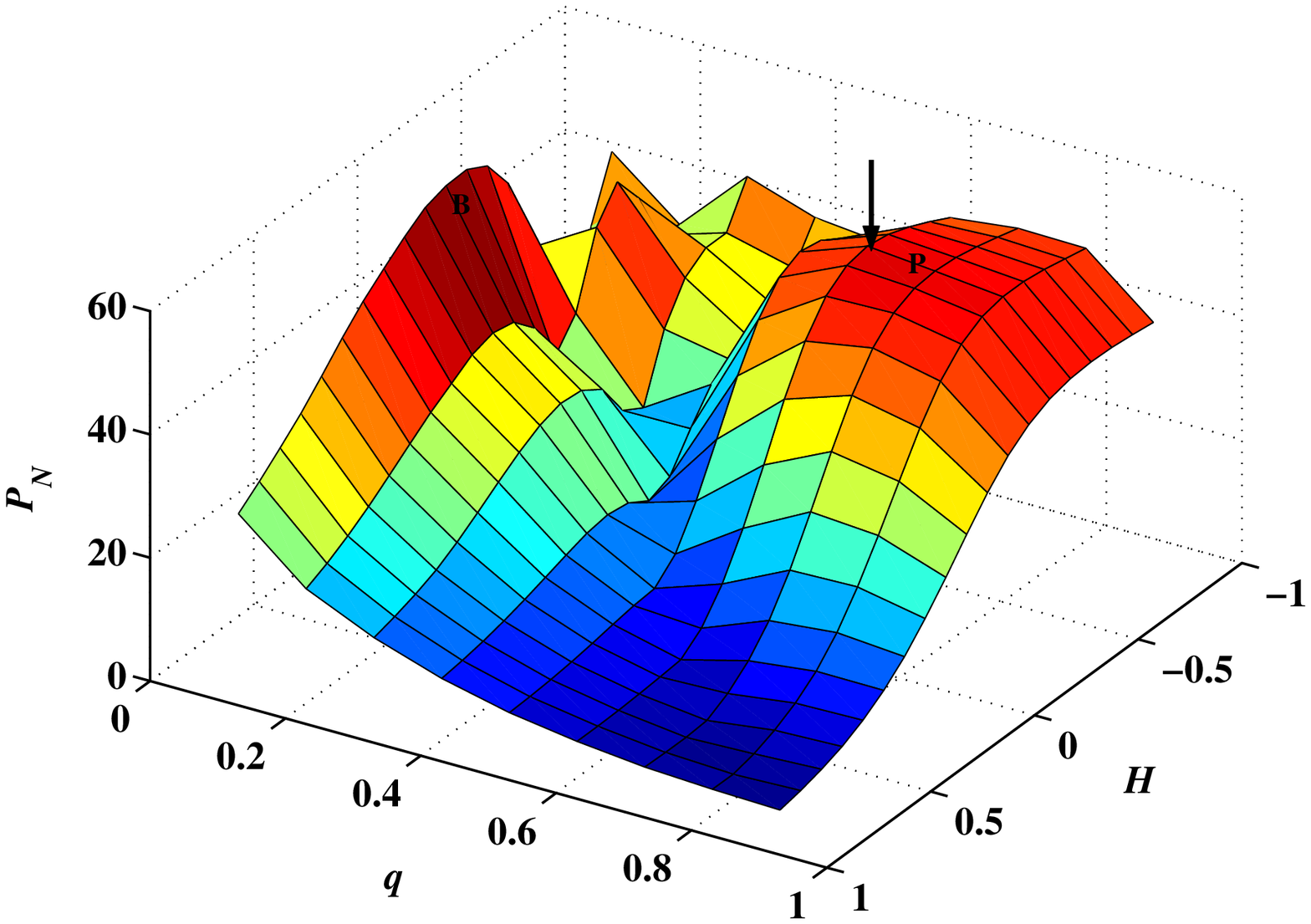,width=11cm, height=8cm}
\end{center}
\caption{Dependence of the height $P_N(H,q)$ of the most
significant peak in each Lomb periodogram of the
$(H,q)$-derivative of the cumulative energy release before the
rupture of tank $\#\,1$.} \label{Fig:DSICumu01PN}
\end{figure}

\begin{figure}
\begin{center}
\epsfig{file=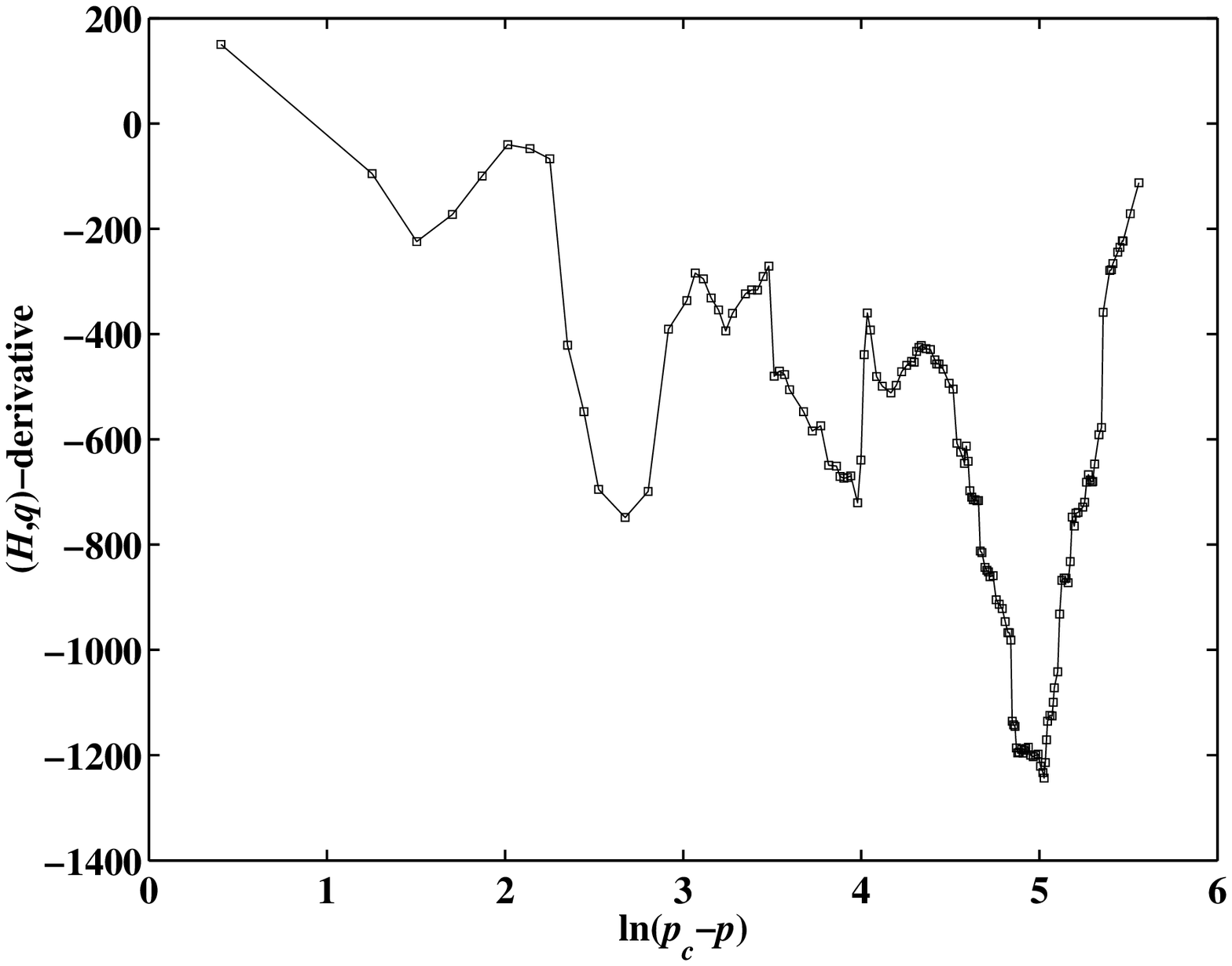,width=12cm, height=9cm}
\end{center}
\caption{$(H,q)$-derivative of the cumulative energy release
before the rupture of tank $\#\,1$ as a function of the
pressure-to-rupture $p_c-p$ with $q=0.6$ and $H=-0.5$.}
\label{Fig:DSICumu01Dy}
\end{figure}

\begin{figure}
\begin{center}
\epsfig{file=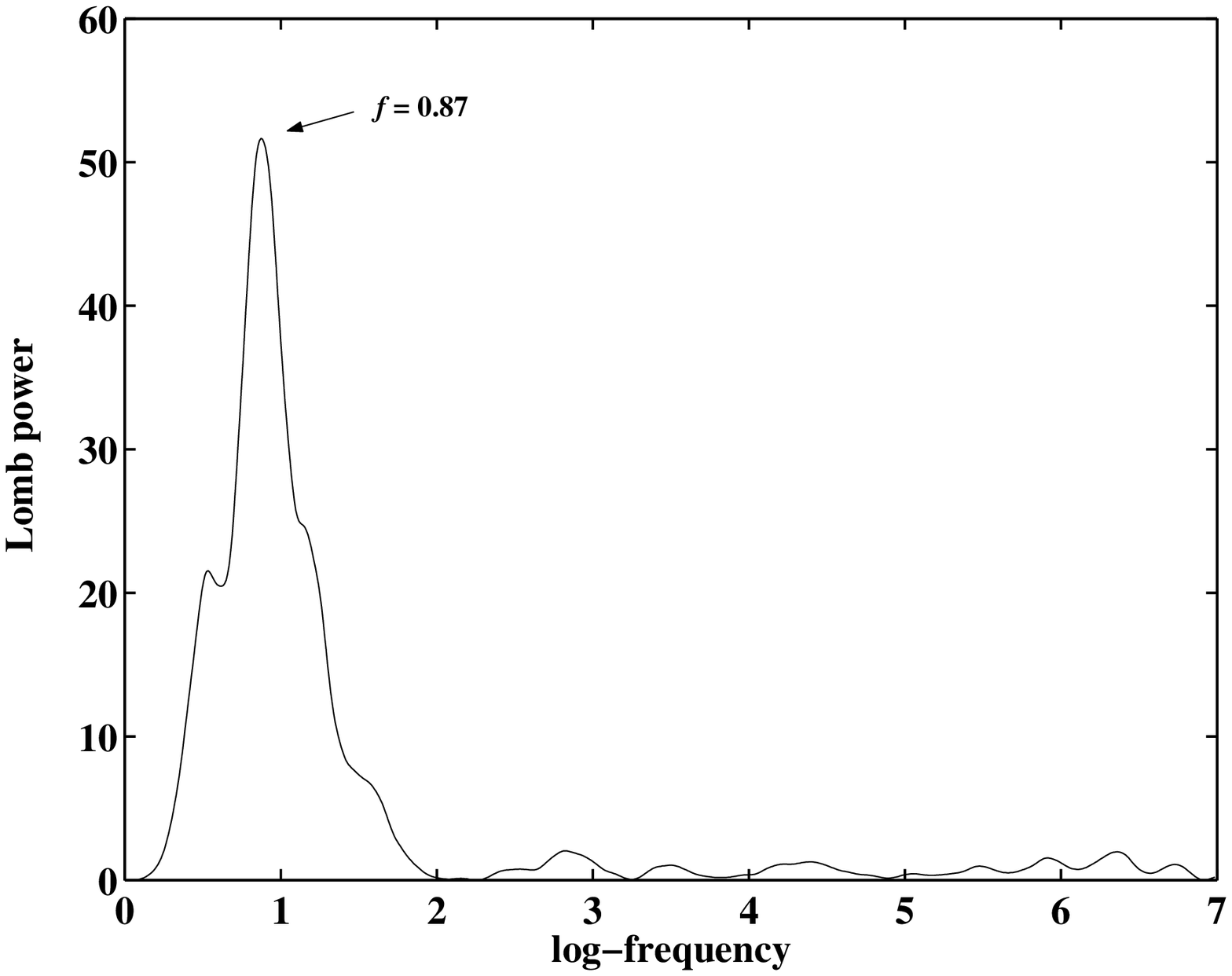,width=12cm, height=9cm}
\end{center}
\caption{Lomb power of the $(H,q)$-derivative shown in
Fig.~\ref{Fig:DSICumu01Dy}.} \label{Fig:DSICumu01Lomb}
\end{figure}

\begin{figure}
\begin{center}
\epsfig{file=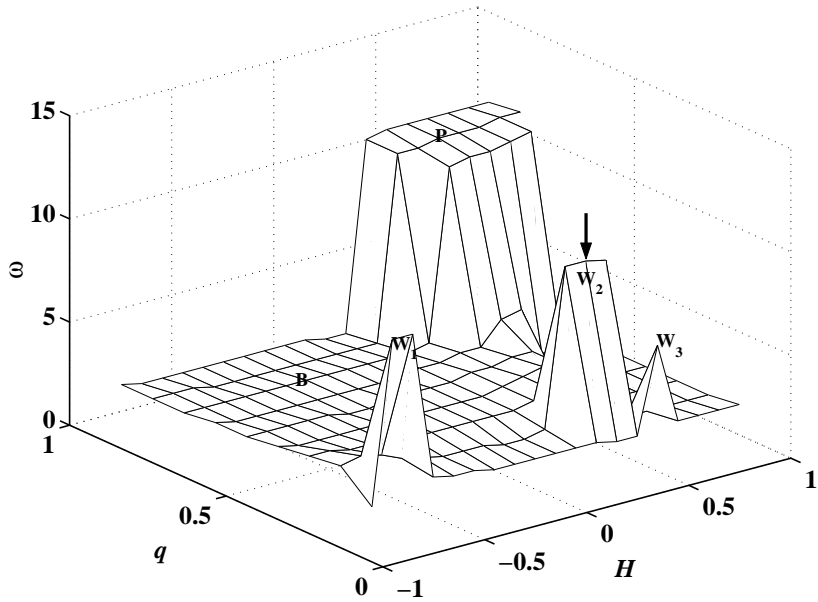,width=11cm, height=8cm}
\end{center}
\caption{Dependence of the logarithmic angular frequency
$\omega(H,q)$ of the most significant peak in each Lomb
periodogram of the $(H,q)$-derivative of the cumulative energy
release before the rupture of tank $\#\,2$.}
\label{Fig:DSICumu02f}
\end{figure}

\begin{figure}
\begin{center}
\epsfig{file=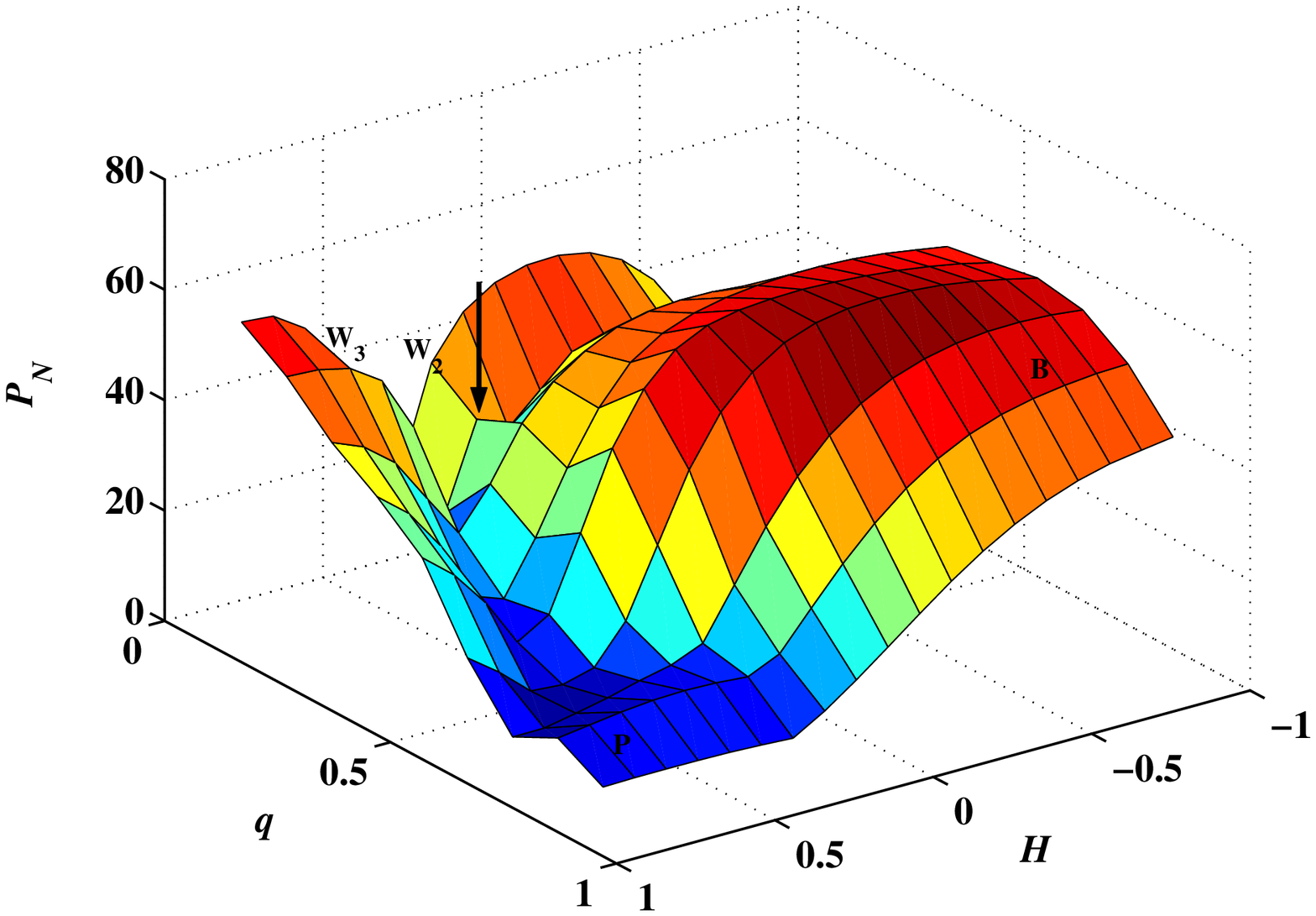,width=11cm, height=8cm}
\end{center}
\caption{Dependence of the height $P_N(H,q)$ of the most
significant peak in each Lomb periodogram of the
$(H,q)$-derivative of the cumulative energy release before the
rupture of tank $\#\,2$.} \label{Fig:DSICumu02PN}
\end{figure}

\begin{figure}
\begin{center}
\epsfig{file=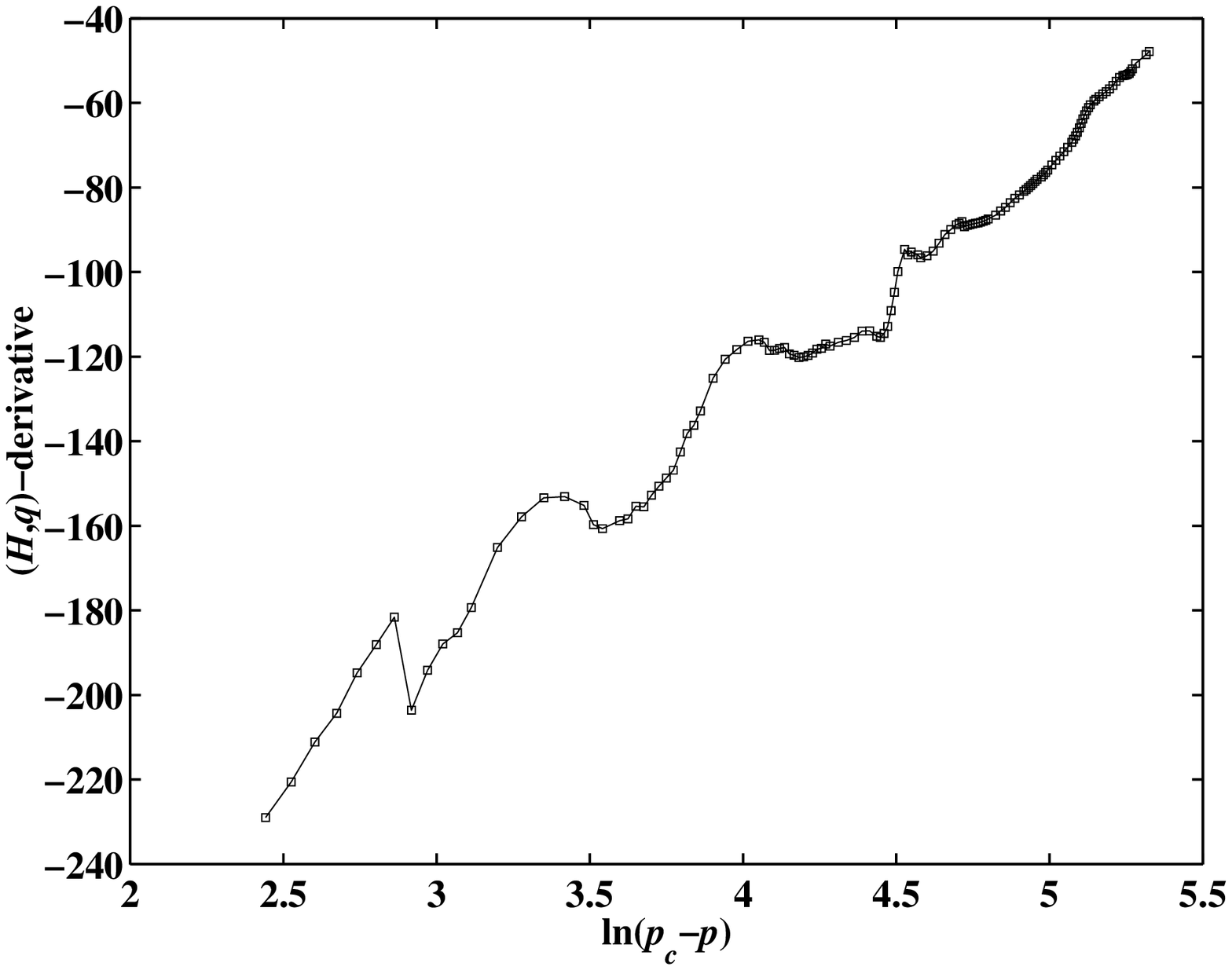,width=12cm, height=9cm}
\end{center}
\caption{$(H,q)$-derivative of the cumulative energy release
before the rupture of tank $\#\,2$ as a function of the
pressure-to-rupture $p_c-p$ with $q=0.2$ and $H=0.3$.}
\label{Fig:DSICumu02Dy}
\end{figure}

\begin{figure}
\begin{center}
\epsfig{file=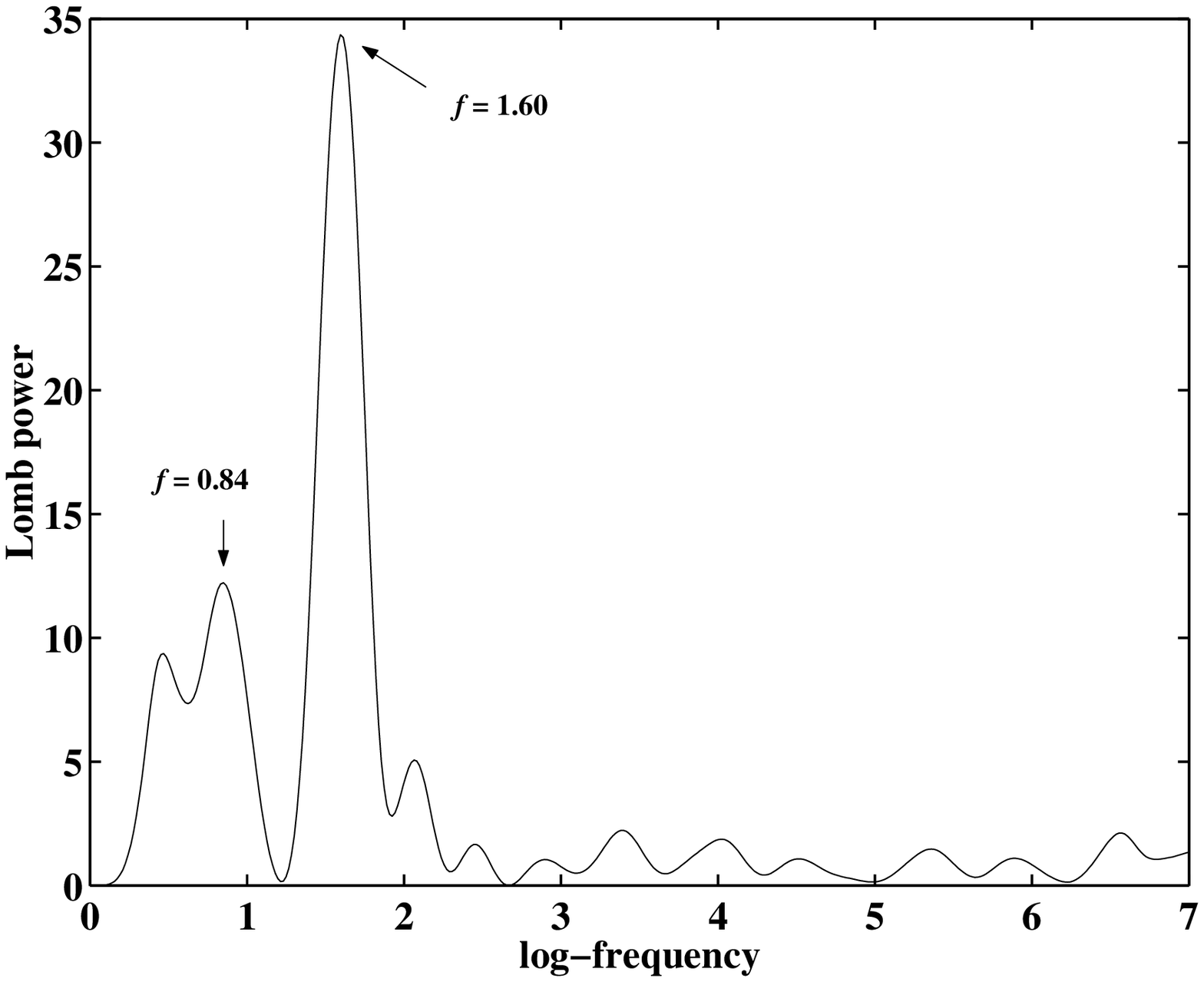,width=12cm, height=9cm}
\end{center}
\caption{Lomb power of the $(H,q)$-derivative shown in
Fig.~\ref{Fig:DSICumu02Dy}. The fundamental log-frequency $f =
0.84$ and its harmonic $f = 1.60$ are indicated by arrows.}
\label{Fig:DSICumu02Lomb}
\end{figure}

\begin{figure}
\begin{center}
\epsfig{file=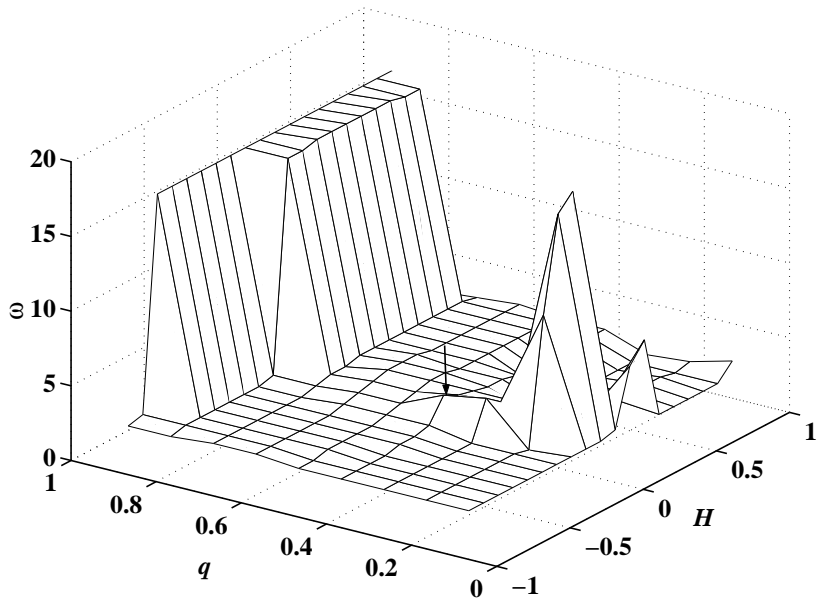,width=11cm, height=8cm}
\end{center}
\caption{Dependence of the logarithmic angular frequency
$\omega(H,q)$ of the most significant peak in each Lomb
periodogram of the $(H,q)$-derivative of the cumulative energy
release before the rupture of tank $\#\,3$. There are two wedges
at $q\sim 0.1$ and $0.2$ excluded according to the fourth criterion,
around which is the bottom of a basin. The platform with $\omega >
15$ is excluded according to third criterion. The optimal pair
$(-0.2, 0.4)$ is located within the second platform $[-0.2,0.9] \times
[0.4,0.7]$.} \label{Fig:DSICumu03f}
\end{figure}

\begin{figure}
\begin{center}
\epsfig{file=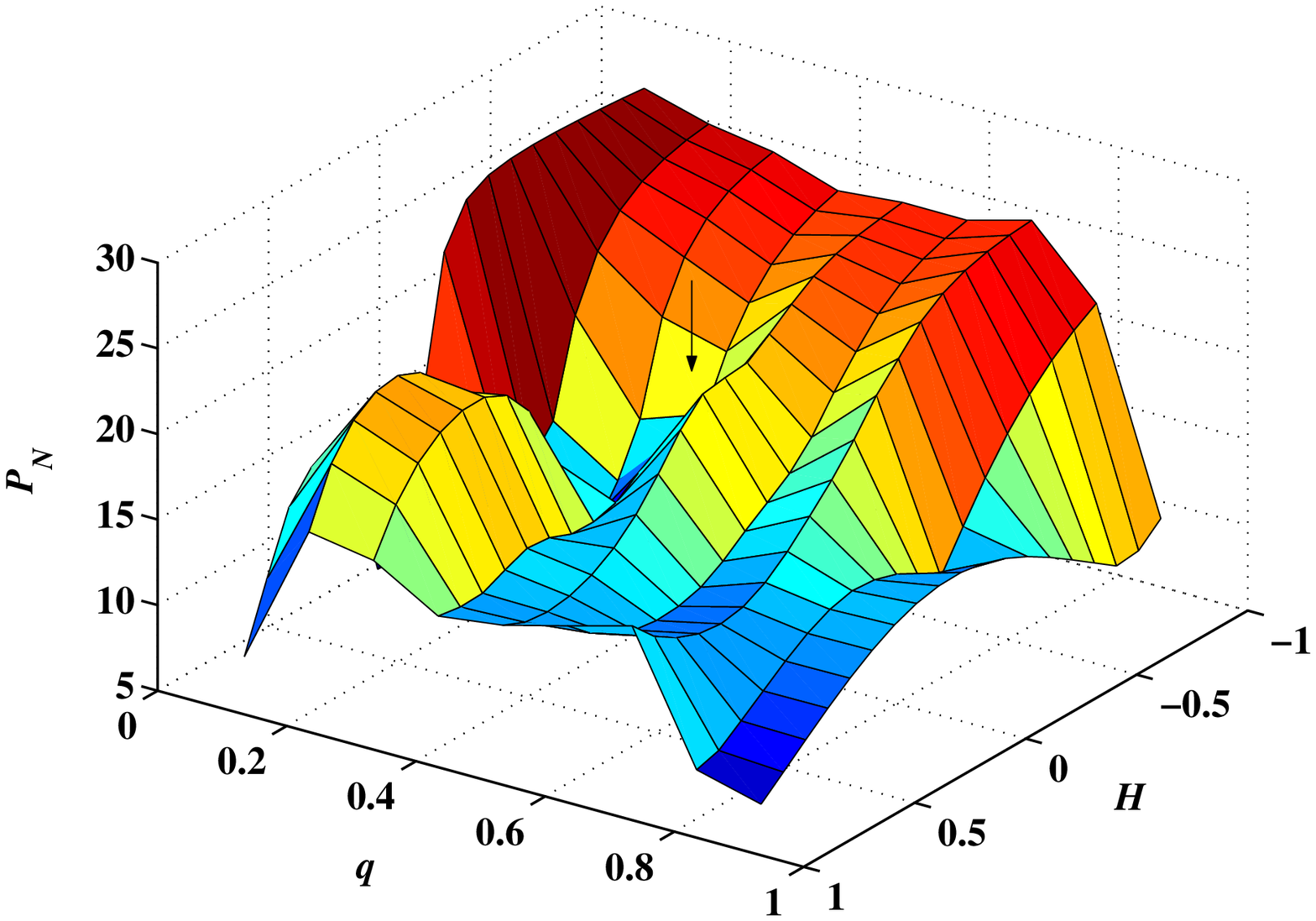,width=11cm, height=8cm}
\end{center}
\caption{Dependence of the height $P_N(H,q)$ of the most
significant peak in each Lomb periodogram of the
$(H,q)$-derivative of the cumulative energy release before the
rupture of tank $\#\,3$.} \label{Fig:DSICumu03PN}
\end{figure}

\begin{figure}
\begin{center}
\epsfig{file=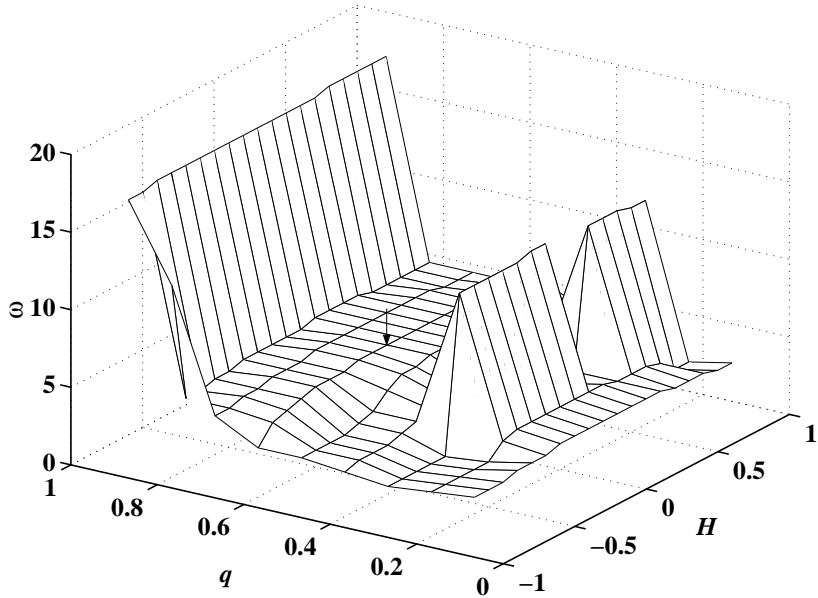,width=11cm, height=8cm}
\end{center}
\caption{Dependence of the logarithmic angular frequency
$\omega(H,q)$ of the most significant peak in each Lomb
periodogram of the $(H,q)$-derivative of the cumulative energy
release before the rupture of tank $\#\,4$. There are two wedges
at $q = 0.3$ excluded according to the fourth criterion. The
``wall'' with $\omega > 15$ is excluded according to third
criterion. The optimal pair $(0, 0.6)$ is located within the platform
indicated by an arrow.} \label{Fig:DSICumu04f}
\end{figure}

\begin{figure}
\begin{center}
\epsfig{file=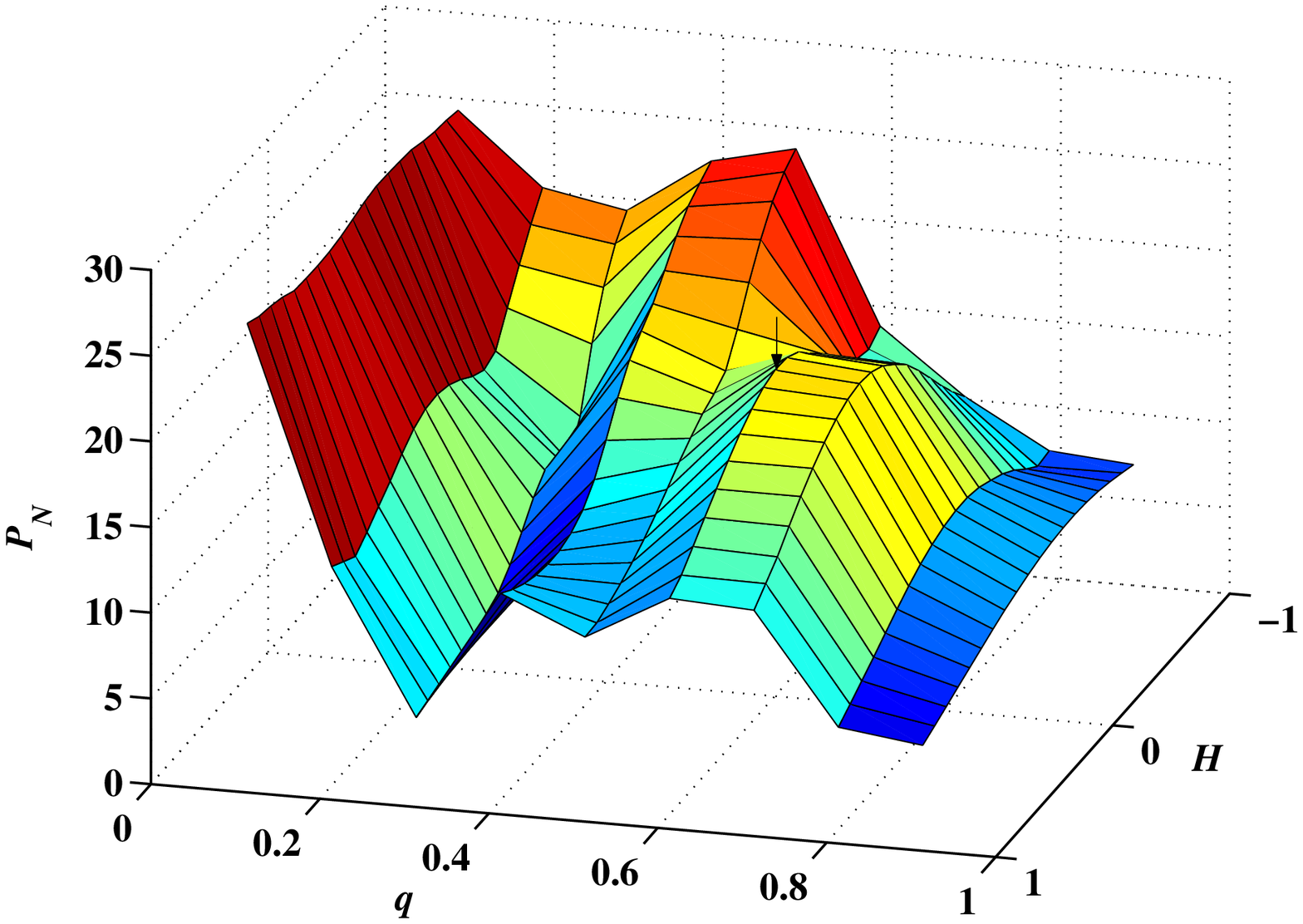,width=11cm, height=8cm}
\end{center}
\caption{Dependence of the height $P_N(H,q)$ of the most
significant peak in each Lomb periodogram of the
$(H,q)$-derivative of the cumulative energy release before the
rupture of tank $\#\,4$. The optimal pair $(0,0.6)$ is indicated
by an arrow.} \label{Fig:DSICumu04PN}
\end{figure}

\begin{figure}
\begin{center}
\epsfig{file=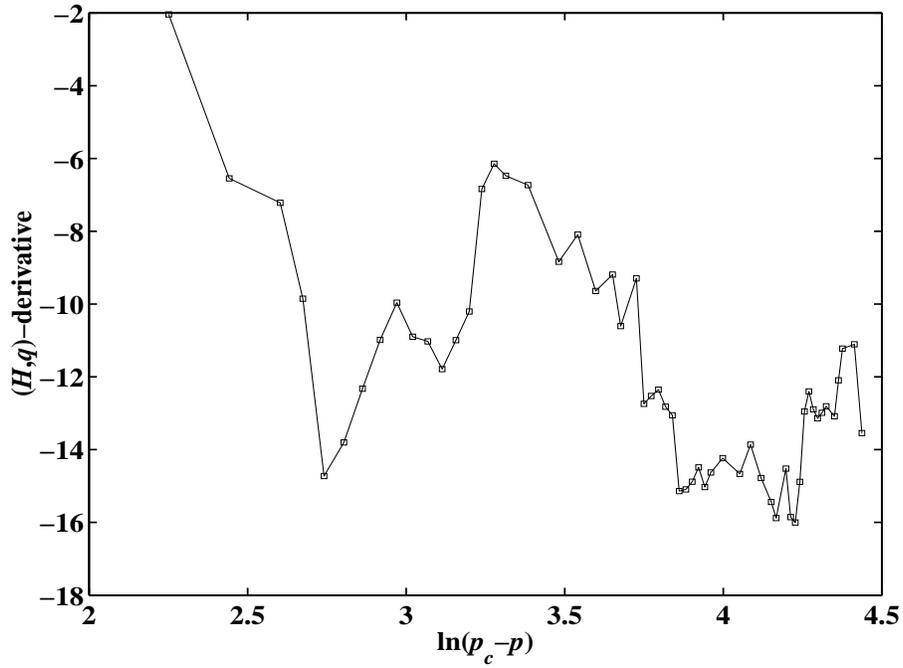,width=12cm, height=9cm}
\end{center}
\caption{$(H,q)$-derivative of the cumulative energy release
before the rupture of tank $\#\,4$ with respect to the
pressure-to-rupture $p_c-p$ with $q=0.6$ and $H=0$.}
\label{Fig:DSICumu04Dy}
\end{figure}

\begin{figure}
\begin{center}
\epsfig{file=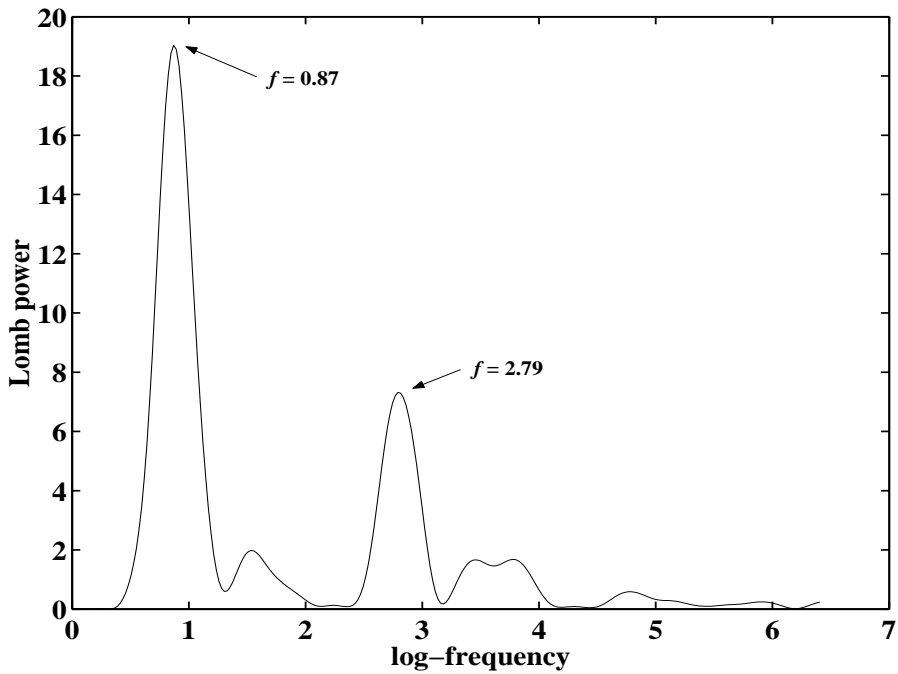,width=12cm, height=9cm}
\end{center}
\caption{Lomb power of the $(H,q)$-derivative shown in
Fig.~\ref{Fig:DSICumu04Dy}. The logarithmic frequencies of the two
highest peaks are $0.87$ and $2.79$, respectively.}
\label{Fig:DSICumu04Lomb}
\end{figure}

\begin{figure}
\begin{center}
\epsfig{file=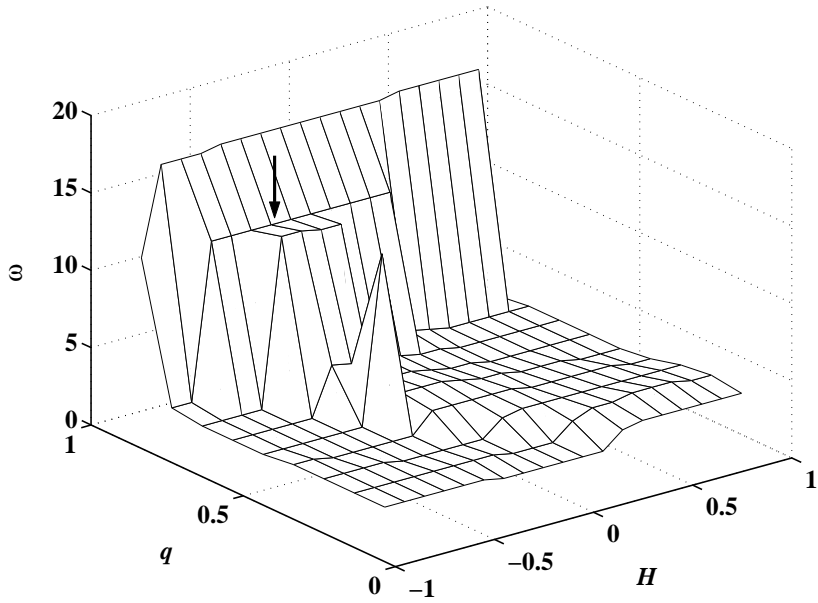,width=11cm, height=8cm}
\end{center}
\caption{Dependence of the logarithmic angular frequency
$\omega(H,q)$ of the most significant peak in each Lomb
periodogram of the $(H,q)$-derivative of the cumulative energy
release before the rupture of tank $\#\,6$. The optimal
pair $(-0.3, 0.8)$ is located within a platform.} \label{Fig:DSICumu06f}
\end{figure}

\begin{figure}
\begin{center}
\epsfig{file=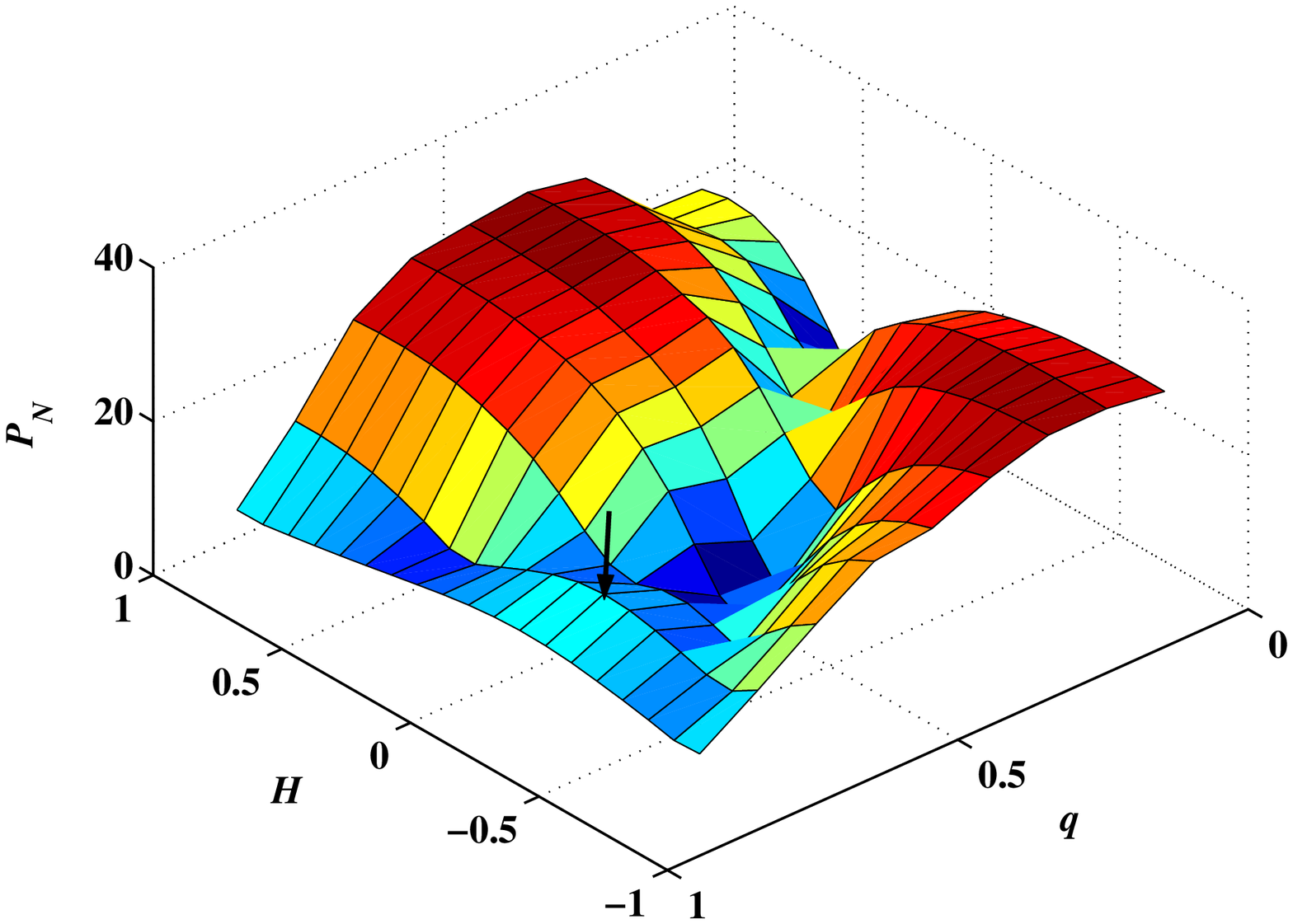,width=11cm, height=8cm}
\end{center}
\caption{Dependence of the height $P_N(H,q)$ of the most
significant peak in each Lomb periodogram of the
$(H,q)$-derivative of the cumulative energy release before the
rupture of tank $\#\,6$. The optimal pair $(-0.3,0.8)$ is
indicated by an arrow.} \label{Fig:DSICumu06PN}
\end{figure}

\begin{figure}
\begin{center}
\epsfig{file=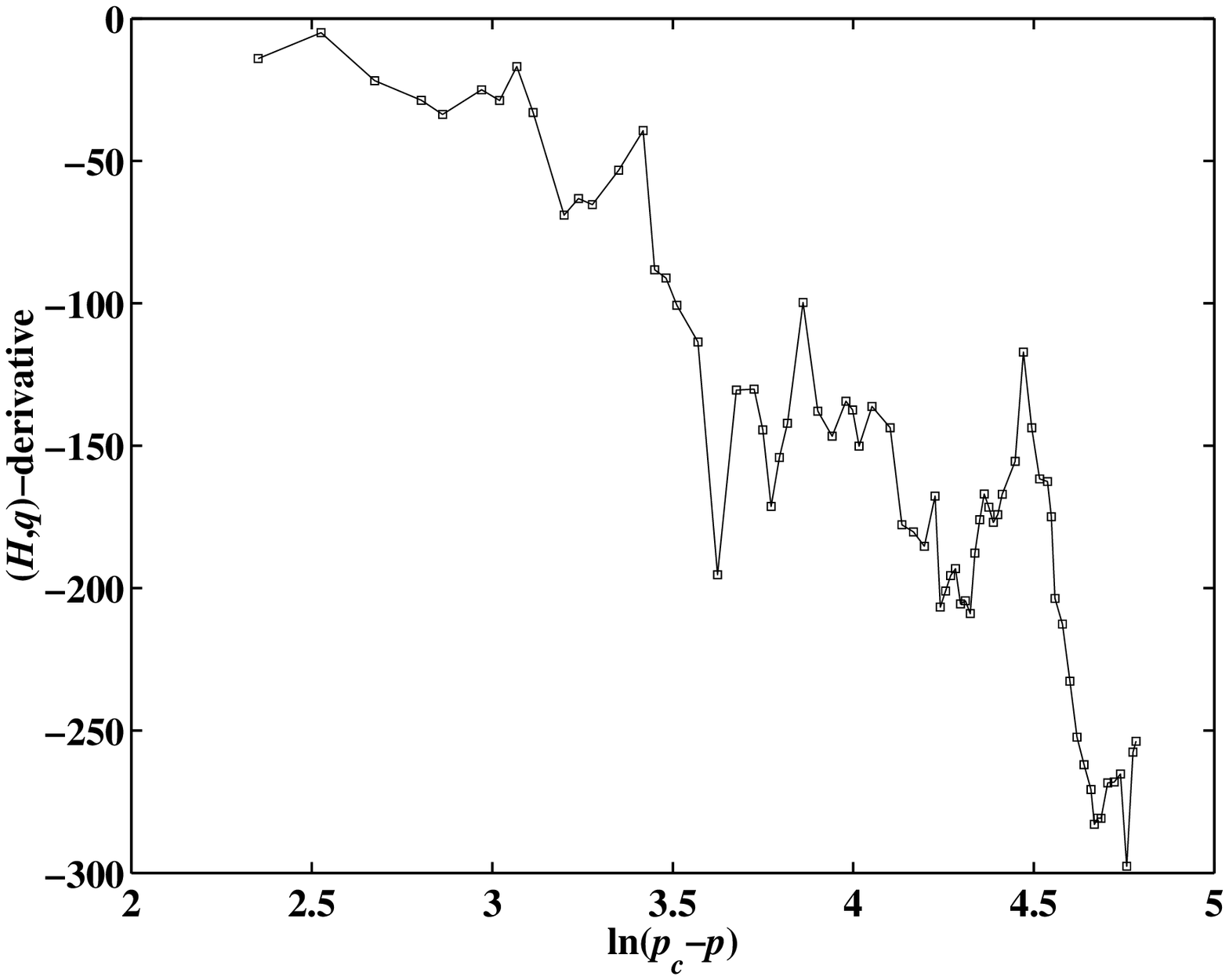,width=12cm, height=9cm}
\end{center}
\caption{$(H,q)$-derivative of the cumulative energy release
before the rupture of tank $\#\,6$ as a function of the
pressure-to-rupture $p_c-p$ with $q=0.8$ and $H=-0.3$.}
\label{Fig:DSICumu06Dy}
\end{figure}

\begin{figure}
\begin{center}
\epsfig{file=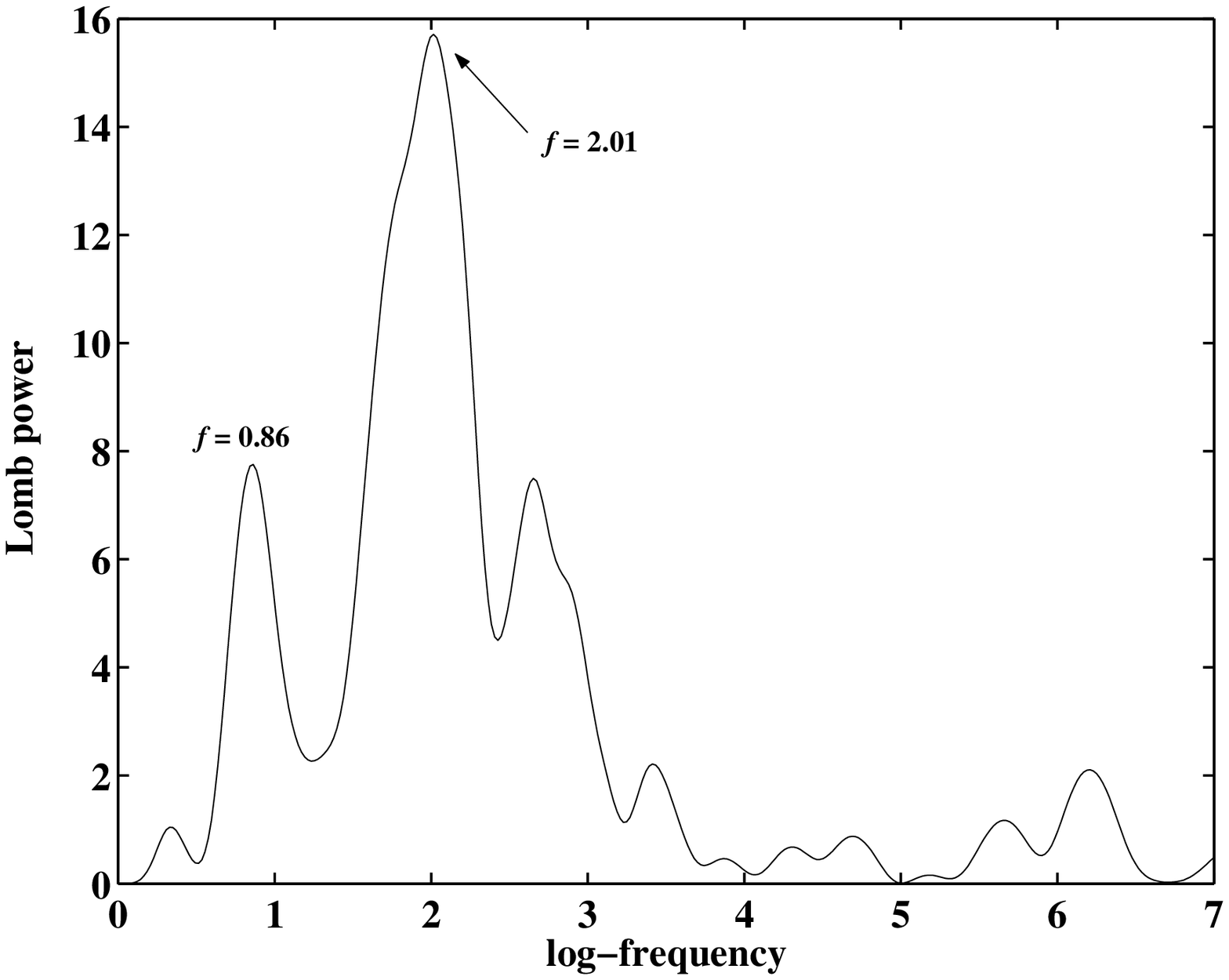,width=12cm, height=9cm}
\end{center}
\caption{Lomb power of the $(H,q)$-derivative shown in
Fig.~\ref{Fig:DSICumu06Dy}.  The fundamental log-frequency $f =
0.86$ and its harmonic $f = 2.01$ are indicated by arrows.}
\label{Fig:DSICumu06Lomb}
\end{figure}

We follow the same procedure as in the previous section to analyze
the cumulative energy release of the same five experimental data
sets discussed previously. We use slightly different truncations
in this analysis of the cumulative energy release. The results
obtained for the five data sets are shown in
Figs.~\ref{Fig:DSICumu01f} to \ref{Fig:DSICumu06Lomb}. Figures
\ref{Fig:DSICumu01PN}, \ref{Fig:DSICumu02PN},
\ref{Fig:DSICumu03PN}, \ref{Fig:DSICumu04PN} and
\ref{Fig:DSICumu06PN} show the dependence of the highest peak
$P_N(H,q)$ in each Lomb periodogram as a function of $H$ and $q$.
Figs.~\ref{Fig:DSICumu01f}, \ref{Fig:DSICumu02f},
\ref{Fig:DSICumu03f}, \ref{Fig:DSICumu04f} and
\ref{Fig:DSICumu06f} are plots of the associated logarithmic
angular frequencies $\omega(H,q)$. The generalized $q$-derivative
of the cumulative energy release corresponding to the optimal
pairs of $(H,q)$ for the five data sets are illustrated
respectively in Figs.~\ref{Fig:DSICumu01Dy},
\ref{Fig:DSICumu02Dy}, \ref{Fig:DSICumu03Dy},
\ref{Fig:DSICumu04Dy} and \ref{Fig:DSICumu06Dy}  with their Lomb
periodograms shown in Figs.~\ref{Fig:DSICumu01Lomb},
\ref{Fig:DSICumu02Lomb}, \ref{Fig:DSICumu03Lomb},
\ref{Fig:DSICumu04Lomb} and \ref{Fig:DSICumu01Lomb}.

In Fig.~\ref{Fig:DSICumu01PN}, the largest $P_N$'s are in the
region ${\mathbf{B}}$. However, the corresponding frequencies are
very low as shown in Fig.~\ref{Fig:DSICumu01f} which gives the
most probable frequency. Hence we excludes region ${\mathbf{B}}$
in accordance with the second criterion. The optimal pair in the
platform ${\mathbf{P}}$ is $(H=-0.5, q=0.6)$. One can observe a
platform of $\omega$ in Fig.~\ref{Fig:DSICumu01f}, which is
consistent with the fourth criterion. In
Fig.~\ref{Fig:DSICumu01Dy}, the log-periodic oscillations of the
$(H,q)$-derivative are clearly visible, and correspond to the high
Lomb peak shown in Fig.~\ref{Fig:DSICumu01Lomb}. The log-periodic
oscillations become even strongly and more significant if more
points are discarded at the low and high pressure ends of the data
set.

Data set $\#\,2$ is more noisy and a major part of the pairs
$(H,q)$ give very low frequencies close to $\omega^{mp}$, as shown
in Fig.~\ref{Fig:DSICumu02f}, which lead us to exclude all the
local maxima shown in Fig.~\ref{Fig:DSICumu02PN}. The optimal pair
$(0.3,0.1)$ is located at the wedge $\mathbf(W_3)$ indicated by an
arrow (in contradiction the fourth criterion). The log-periodicity
is quite apparent in Fig.~\ref{Fig:DSICumu02Dy}. Correspondingly,
its Lomb peak shown in Fig.~\ref{Fig:DSICumu02Lomb} is very
significant. The data sets $\#\,3$ and $\#\,4$ are similar to
$\#\,2$. For $\#\,6$, the optimal pair $(-0.3,0.8)$ located on the
platform shown in Fig.~\ref{Fig:DSICumu06f} has a local maximum
shown in Fig.~\ref{Fig:DSICumu06PN}. We summarize all these
results in Table \ref{Tb:cumu}.

\begin{table}
\begin{center}
\begin{tabular}{|c|c|c|c|c|c|c|c|c|c|c|c|c|}
\hline Tank&Points&$p_{\min}$&$p_{\max}$&$p_c$&$(H,q)$&$P_N$
&$\omega$&$\omega '$&$\omega^{mp}$&$N_{osc}$&Gaussian&fGn\\\hline
$\#\,1$&144&453.5&711.5&713&$(-0.5,0.6)$&51.7&5.5&$/$&1.8&4.6&5E-21&0.
98\\\hline
$\#\,2$&138&467.5&661.5&673&$(0.3,0.2)$&34.4&10.0&5.3&3.3&4.5&2E-13&0.
93\\\hline
$\#\,3$&63&671.5&756.5&764&$(-0.2,0.4)$&14.5&4.5&9.9&3.8&1.8&5E-5&0.70\\\hline
$\#\,4$&56&671.5&746.5&756&$(0,0.6)$&19.0&5.5&17.5&4.3&1.9&3E-7&0.79\\\hline
$\#\,6$&72&614.5&723.5&734&$(-0.3,0.8)$&15.7&12.7&5.4&3.9&4.9&1E-5&0.71\\\hline
\end{tabular}
\caption{Summary of the results of the $(H,q)$-analysis on the
cumulative energy release of data sets $\#\,1$, $\#\,2$, $\#\,3$,
$\#\,4$ and $\#\,6$. The meaning of the columns are the same as
in Table \ref{Tb:form}.} \label{Tb:cumu}
\end{center}
\end{table}

\subsection{Comparison between energy release rates and their
cumulative} \label{ss:cmp}

\begin{figure}
\begin{center}
\epsfig{file=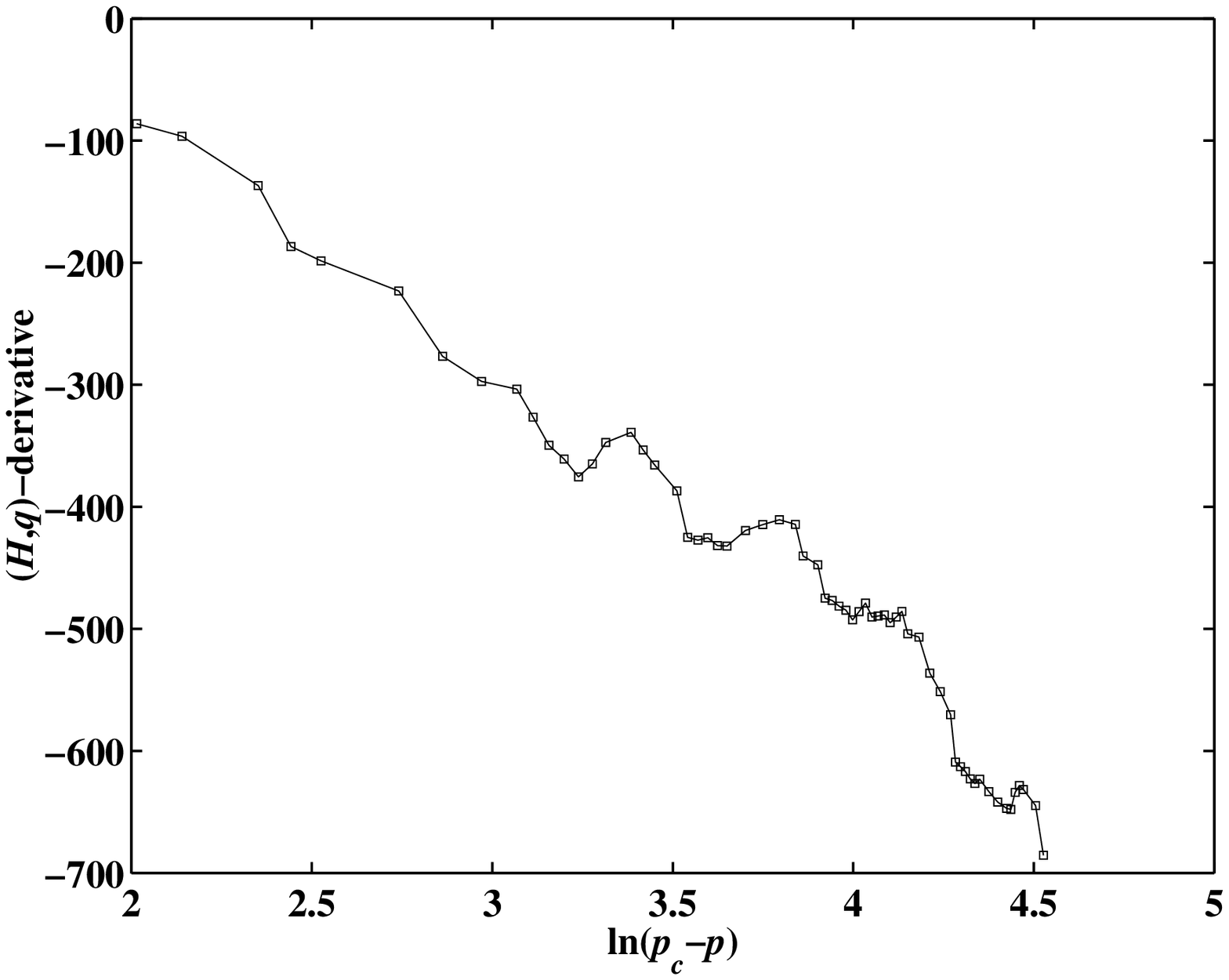,width=12cm, height=9cm}
\end{center}
\caption{$(H,q)$-derivative of the cumulative energy release
before the rupture of tank $\#\,3$ as a function of the
pressure-to-rupture $p_c-p$ with $q=0.4$ and $H=-0.2$.}
\label{Fig:DSICumu03Dy}
\end{figure}

\begin{figure}
\begin{center}
\epsfig{file=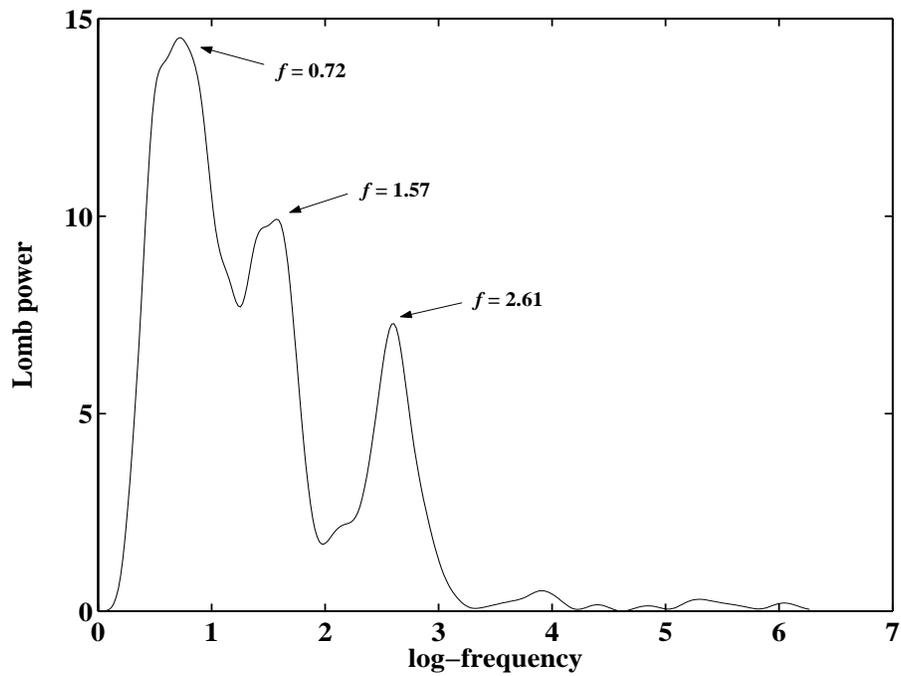,width=12cm, height=9cm}
\end{center}
\caption{Lomb power of the $(H,q)$-derivative shown in
Fig.~\ref{Fig:DSICumu03Dy}. The logarithmic frequencies of the
three highest peaks are $0.72, 1.57$ and $2.61$, respectively.}
\label{Fig:DSICumu03Lomb}
\end{figure}

Comparing Table \ref{Tb:form} and Table \ref{Tb:cumu}, we find
that the cumulative energy releases present a more significant
log-periodic component. Taking the cumulative corresponds to
perform a low-pass filter that smooths out significantly the noise
and usually provides better signals with higher signal-to-noise
ratio. Of course, taking the cumulative also reduces the amplitude
of the underlying log-periodic oscillations. There is thus a
compromise. As an illustration, let us compare
Fig.~\ref{Fig:DSIForm03Dy} and Fig.~\ref{Fig:DSICumu03Dy}. It is
clear that the $(H,q)$-derivative of the energy release rate in
Fig.~\ref{Fig:DSIForm03Dy} is much more noisy than that of the
cumulative energy release shown in Fig.~\ref{Fig:DSICumu03Dy}. On
the other hand, the high-frequency oscillations which are clearly
visible in Fig.~\ref{Fig:DSICumu03Dy} are suppressed in
Fig.~\ref{Fig:DSICumu03Dy} and can not be captured by the Lomb
analysis. Thus, the log-frequency with largest peak power of the
energy rate ($f=1.97$) is much higher than that of the cumulative
energy ($f=0.72$). We note however that the second highest peak in
Fig.~\ref{Fig:DSICumu03Dy} corresponds to the highest peak of the
energy release rate.

In general, we find that the cumulative energy release keeps the
same underlying log-periodic structure as found in the energy
release rate. We observe that data sets $\#\,1$, $\#\,2$ and
$\#\,4$ have very similar logarithmic frequencies for the energy
release rate and their cumulative. Most Lomb periodograms exhibit
harmonics of each other. This close correspondence obtained in the
$(H,q)$-analysis provides a better characterization of the
log-periodic nature in the system than obtained previously with
the parameterization approach of fitting the data sets with a
log-periodic function \cite{Critrup}.

Finally, we note that all data sets exhibit a fundamental value of
the logarithmic angular frequency which is compatible with the
value $\omega = 5.8\pm 0.9$. In some cases, it is its second
harmonic that gives the strongest peak of the Lomb periodograms.
This value corresponds to a preferred scaling ratio $\lambda =
e^{2 \pi/\omega} = 3.0 \pm 0.5$.

\section{Post-diction of ruptures}
\label{s:post}

The $(H,q)$-analysis has shown its power for detecting
log-periodicity, conditioned on our knowledge of the critical
pressure $p_c$ at rupture. It is natural to ask whether it can be
extended to provide advanced prediction of $p_c$. For carrying
these tests, we use the cumulative energy release which provided
the strongest log-periodic signal in the analysis reported in
previous sections. Our strategy is to use each data up to a
maximum pressure $p_{\max} < p_c$, assume some value for $p_c$ and
perform the same analysis as in section 3. i.e., for a given
presumed critical points $p_c$, we determine the optimal pair
$(H,q)$. For each presumed $p_c$, we thus obtain the Lomb peak
height $P_N(H,q)$ and its associated logarithmic angular frequency
$\omega(H,q)$ as a function of $H$ and $q$. In order to find the
optimal $(H,q)$ corresponding to the maximum $P_N(H,q)$, we add
the second criterion to exclude those $(H,q)$ with $\omega(H,q)
\leq \omega^{mp}$. Having determined the optimal $P_N(p_c)$ and
$\omega(p_c)$ as functions of the presumed critical pressure
$p_c$, we determine the predicted critical pressure $\hat{p}_c$ by
the condition \be P_N(\hat{p}_c) = \max_{p_c} \{P_N(p_c)\},
\label{Eq:PNpc} \ee without further constraints on $\omega(p_c)$.

\begin{figure}
\begin{center}
\epsfig{file=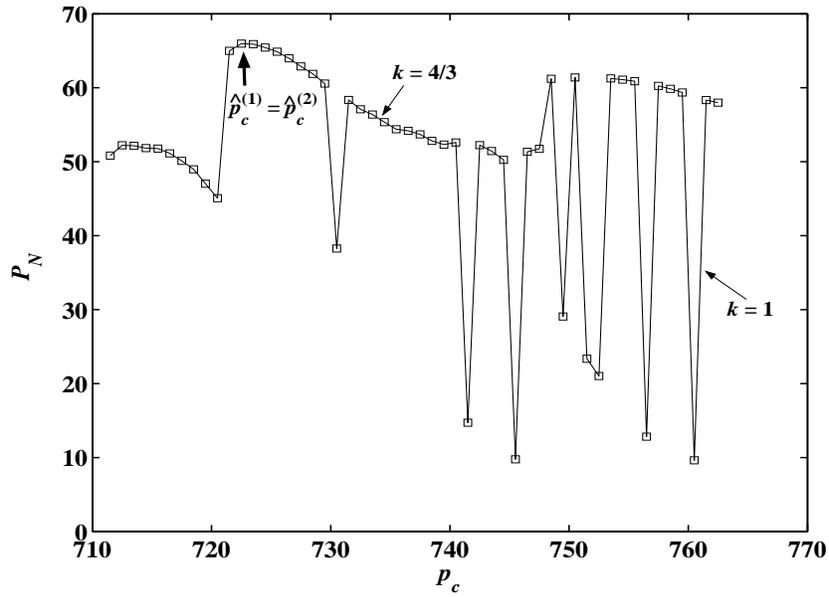,width=11cm, height=8cm}
\end{center}
\caption{Post-diction of the critical pressure of the rupture
of tank $\#\,1$. The two fine arrows indicate the upper threshold
of predictable $\hat{p}_c$ estimated by Eq.~(\ref{Eq:upbound2})
with $k=4/3$ and $k=1$. The coarse arrows indicate the predicted
critical pressures $\hat{p}_c^{(1)}$, $\hat{p}_c^{(2)}$ and
$\hat{p}_c^{(3)}$.} \label{Fig:Post01}
\end{figure}

\begin{figure}
\begin{center}
\epsfig{file=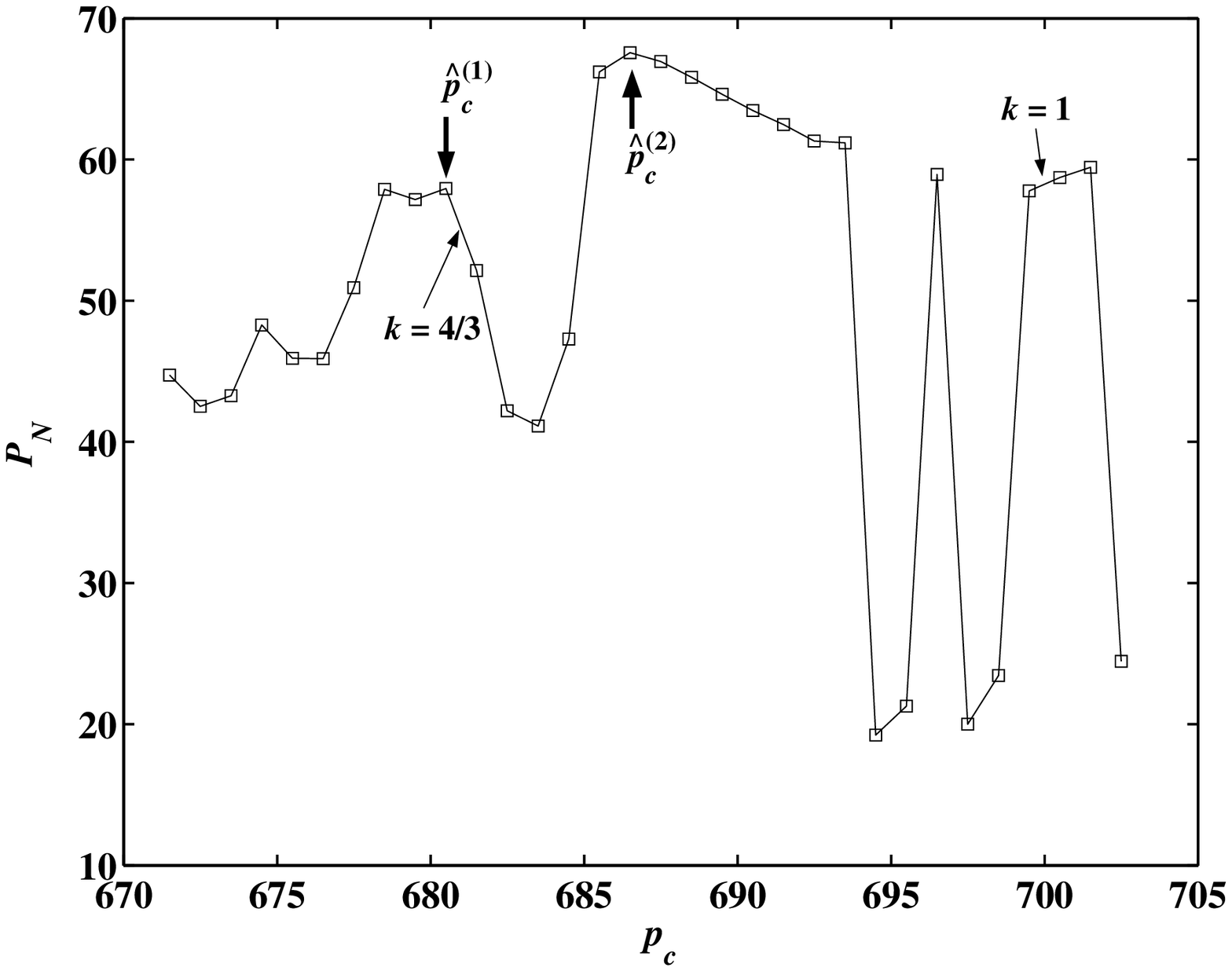,width=11cm, height=8cm}
\end{center}
\caption{Post-diction of the critical pressure of the rupture
of tank $\#\,2$.  Same as in figure \ref{Fig:Post01}.} \label{Fig:Post02}
\end{figure}

\begin{figure}
\begin{center}
\epsfig{file=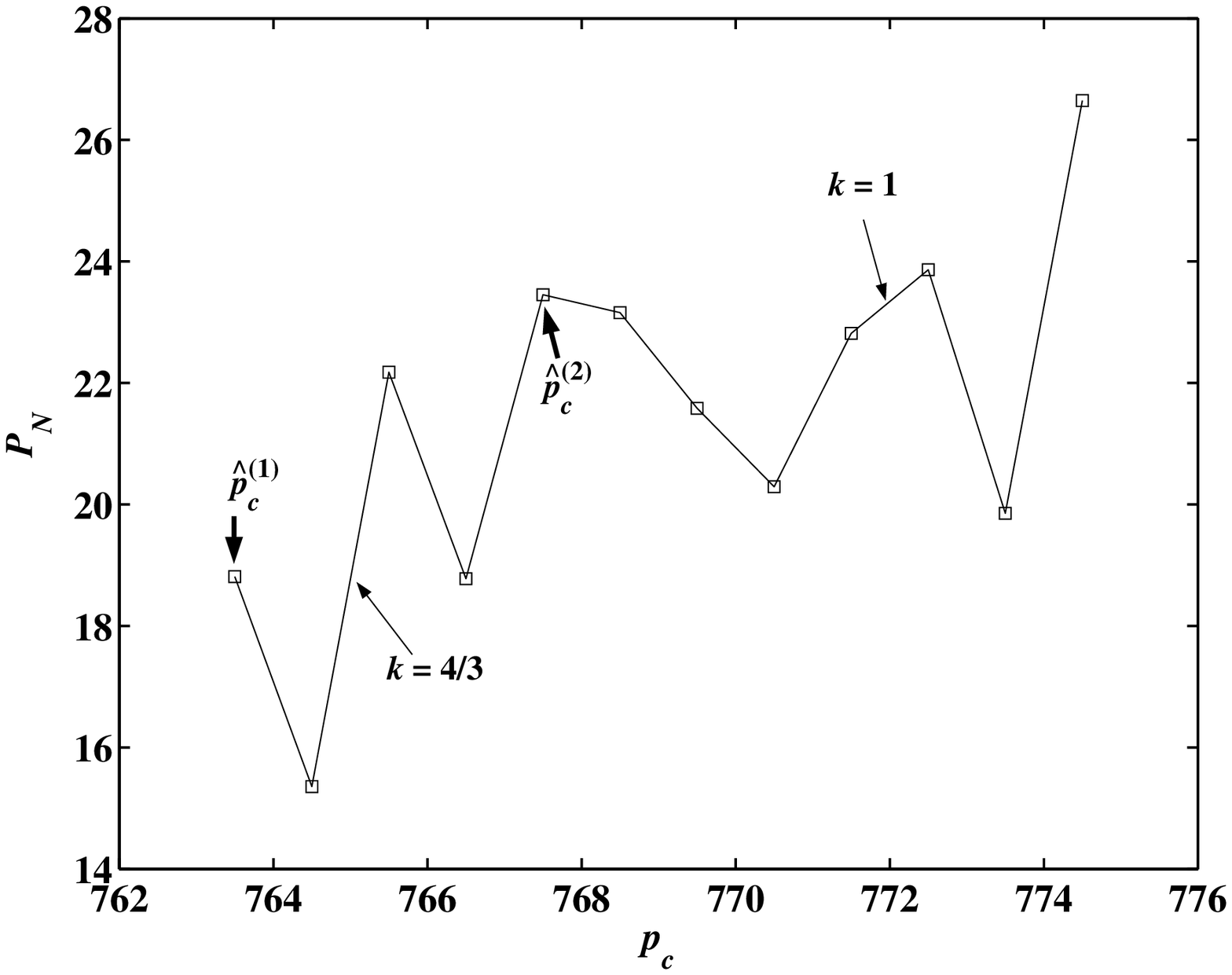,width=11cm, height=8cm}
\end{center}
\caption{Post-diction of the critical pressure of the rupture
of tank $\#\,3$. Same as in figure \ref{Fig:Post01}.} \label{Fig:Post03}
\end{figure}

\begin{figure}
\begin{center}
\epsfig{file=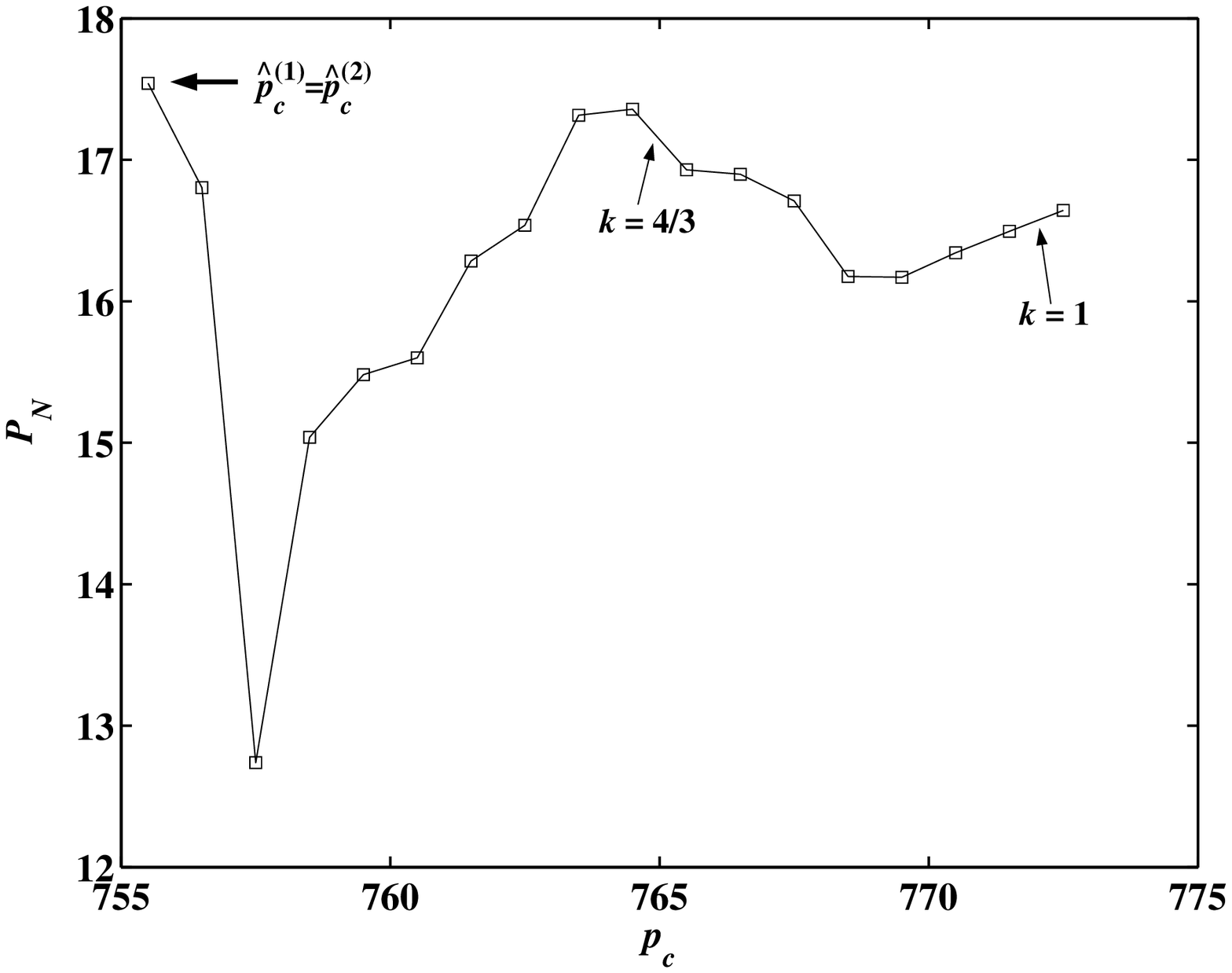,width=11cm, height=8cm}
\end{center}
\caption{Post-diction of the critical pressure of the rupture
of tank $\#\,4$. Same as in figure \ref{Fig:Post01}.} \label{Fig:Post04}
\end{figure}

\begin{figure}
\begin{center}
\epsfig{file=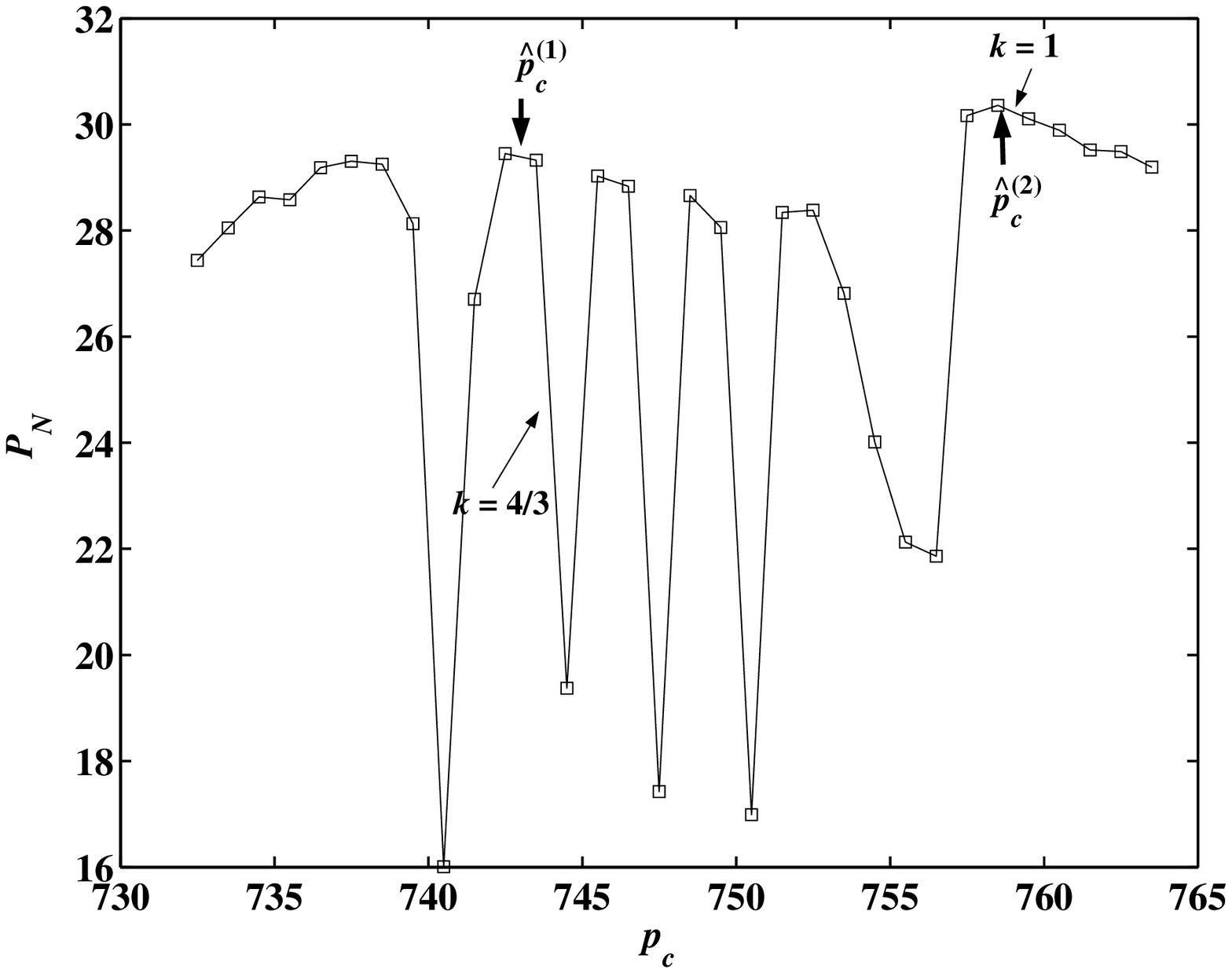,width=11cm, height=8cm}
\end{center}
\caption{Post-diction of the critical pressure of the rupture
of tank $\#\,6$. Same as in figure \ref{Fig:Post01}.} \label{Fig:Post06}
\end{figure}

In practice, we analyze the function $P_N(p_c; H,q)$ of the three
variables $p_c, H$ and $q$ in the following sequence: we fix a value for
$p_c$ and explore the plane $(H,q)$. We then change $p_c$ and redo
the exploration of the plane $(H,q)$, etc.
Fig.~\ref{Fig:Post01} through
\ref{Fig:Post06} give the optimal Lomb peak height $P_N$ as a function
of the presumed critical pressure $p_c$. It is obvious that this
scheme does not give a good predictive skill, as the most important
pattern observed in these figures is the systematic increasing trend,
tending to reject the predicted $p_c$ towards too large values.

We thus attempt to improve this prediction skill by requesting
that the predicted $p_c$ should not be too far from the last
point, i.e., it is nonsensical to hope to predict too far from the
``present.'' To implement this idea, we impose $\omega >
k\omega^{mp}$, namely, \be {\omega \over 2\pi} > {1.5k \over
\ln(p_c-p_{\min}) -\ln(p_c-p_{\max})}, \label{Eq:upbound1} \ee
where $k\geq 1$ is a ``safety factor.'' This constraint translates
into the following condition for $p_c$: \be p_c < p_{\max} +
{p_{\max} - p_{\min} \over e^{3k\pi \over \omega} - 1}.
\label{Eq:upbound2} \ee This constraint (\ref{Eq:upbound2}) means
that there exists an upper threshold beyond which we can not make
a physically meaningful prediction. Since the left-hand-side of
(\ref{Eq:upbound2}) is monotonously increasing with $\omega$, it
is possible to make a prediction much earlier before the critical
point, the larger $\omega$ is. This is natural since large
$\omega$ implies more log-periodic oscillations and thus a
stronger log-periodic signal. To implement this condition in
practice, we take $\omega = 5.8$ as the central value of the
log-periodic angular frequency. According to this constraint
(\ref{Eq:upbound2}), ruptures of tanks $\#\,5$ and $\#\,7$ are
unpredictable, since the last point $p_{\mathtt{last}}$ is too far
from the true critical pressure.

The results are given in Table \ref{Tb:post}. The columns
$\hat{p}_c^{(1)}$ and $\hat{p}_c^{(2)}$ list the predicted
critical pressure $\hat{p}_c$ and its corresponding upper
threshold in the parenthesis with the constraint
(\ref{Eq:upbound2}) for the two choices $k=4/3$ and $k=1$
respectively. The prediction for experiment $\#\,1$ is very good.
For the other experiments, $\hat{p}_c$ is found to be very close
to the upper threshold. The predictions for experiments $\#\,2$,
$\#\,3$ and $\#\,4$ are reasonable while the prediction for case
$\#\,6$ fails completely.

\begin{table}
\begin{center}
\begin{tabular}{|c|c|c|c|c|c|c|c|c|}
\hline
Tank&Points&$p_{\min}$&$p_{\max}$&$\omega$&$p_c$&$p_c-p_{\max}
\over p_c$ &$\hat{p}_c^{(1)}$
&$\hat{p}_c^{(2)}$ \\\hline
$\#\,1$&139&453.5&705.5&5.5&713&$1.1\%$&722.5(734)&722.5(761)\\\hline
$\#\,2$&149&452.5&662.5&5&673&$1.6\%$&680.5(681)&686.5(700)\\\hline
$\#\,3$&65&671.5&759.5&4.5&764&$0.6\%$&763.5(765)&767.5(772)\\\hline
$\#\,4$&56&671.5&746.5&5.5&756&$1.3\%$&755.5(755)&755.5(763)\\\hline
$\#\,6$&74&614.5&726.5&6.4&734&$1.0\%$&742.5(744)&758.5(759)\\\hline
\end{tabular}
\caption{Post-diction using the $(H,q)$-analysis on the cumulative
energy release of tanks $\#\,1$, $\#\,2$, $\#\,3$, $\#\,4$ and
$\#\,6$. Columns $\hat{p}_c^{(1)}$ and $\hat{p}_c^{(2)}$ list the
predicted critical pressure $\hat{p}_c$ and its corresponding
upper threshold in the parenthesis with the constraint
(\ref{Eq:upbound2}) for $k=4/3$ and $k=1$ respectively. The
meaning of the other columns are the same as in Table
\ref{Tb:form}.} \label{Tb:post}
\end{center}
\end{table}

\bigskip
{\bf Acknowledgments:} We are grateful to A. Erzan for a
discussion on Jackson's integral and for supplying the
corresponding references. This work was supported by
NSF-DMR99-71475 and the James S. Mc Donnell Foundation 21st
century scientist award/studying complex system.

\pagebreak

\section*{APPENDIX A}

 From the definition (\ref{Eq:bracket}), the extremal condition
$dg(x)/dx = 0$ gives the extreme points of
$g(x)$ solutions of
\begin{equation}
C_1 \sin(\omega \ln{x}) - C_2\cos(\omega\ln{x}) = 0~,
\label{Eq:d1g}
\end{equation}
that is,
\begin{equation}
\omega \ln(x_m) = n\pi + \arctan(C_2/C_1), \label{Eq:xm}
\end{equation}
where $n = 0,\pm 1,\pm 2,\cdots$. It follows that the extreme
values of $g(x)$ are
\begin{equation}
g(x_m) = \pm \sqrt{C_1^2+C_2^2}~. \label{Eq:gm}
\end{equation}
When $\omega \ln(x_m) = 2n\pi + \arctan(C_2/C_1)$, $g(x_m) =
\sqrt{C_1^2+C_2^2}$ is the maximum; when $\omega \ln(x_m) =
(2n+1)\pi + \arctan(C_2/C_1)$, $g(x_m) = -\sqrt{C_1^2+C_2^2}$ is
the minimum. This is proved by the following exhaustive
classification.

\begin{enumerate}
\item
$e^{(2k-1)\pi/\omega}< q < e^{2k\pi/\omega}$, with $k = 0,\pm
1,\pm 2,\cdots$.

In this case, $C_2 < 0$. Thus $-\pi/2 < \arctan(C_2/C_1) < 0$.
When $\omega \ln(x_m) = 2n\pi + \arctan(C_2/C_1)$, $\cos(\omega
\ln{x_m})>0$ and $\sin(\omega \ln{x_m})<0$, and thus $g(x_m) =
\sqrt{C_1^2+C_2^2}$ is the maximum. When $\omega \ln(x_m) =
(2n+1)\pi + \arctan(C_2/C_1)$, $\cos(\omega \ln{x_m})<0$ and
$\sin(\omega \ln{x_m})>0$, and thus $g(x_m) = -\sqrt{C_1^2+C_2^2}$
is the minimum.

\item
$q = e^{k\pi/\omega}$, with $k = 0,\pm 1,\pm 2,\cdots$.

In this case, $C_2=0$. Thus $\arctan(C_2/C_1) = 0$ and $g(x) =
C_1\cos(\omega\ln{x})$. When $\omega \ln(x_m) = 2n\pi$, $g(x_m) =
C_1$ is the maximum. When $\omega \ln(x_m) = (2n+1)\pi$, $g(x_m) =
-C_1$ is the minimum.

\item
$e^{2k\pi/\omega}< q < e^{(2k+1)\pi/\omega}$, with $k = 0,\pm
1,\pm 2,\cdots$.

In this case, $C_2>0$. Thus $0 < \arctan(C_2/C_1) < \pi/2$. When
$\omega \ln(x_m) = 2n\pi + \arctan(C_2/C_1)$, $\cos(\omega
\ln{x_m})>0$ and $\sin(\omega \ln{x_m})>0$, and thus $g(x_m) =
\sqrt{C_1^2+C_2^2}$ is the maximum. When $\omega \ln(x_m) =
(2n+1)\pi + \arctan(C_2/C_1)$, $\cos(\omega \ln{x_m})<0$ and
$\sin(\omega \ln{x_m})<0$, and thus $g(x_m) = -\sqrt{C_1^2+C_2^2}$
is the minimum.
\end{enumerate}
The amplitude of $D_q^H y(x)$ is then
\begin{equation}
A = C'\sqrt{C_1^2+C_2^2}~(x_m)^{m-H}, \label{Eq:Amp2}
\end{equation}
while the local extreme values are
\begin{equation}
D_m = \left(B' \pm C'\sqrt{C_1^2+C_2^2}\right)(x_m)^{m-H}.
\label{Eq:Dm2}
\end{equation}

\pagebreak

%FIGURE 1
\clearpage

%FIGURE 2

\clearpage

%FIGURE 3

%FIGURE 4

%FIGURE 5

%FIGURE 6

\clearpage
%FIGURE 7

%FIGURE 8

%FIGURE 9

%FIGURE 10

\clearpage
%FIGURE 11

%FIGURE 12

%FIGURE 13

%FIGURE 14

\clearpage
%FIGURE 15

%FIGURE 16

%FIGURE 17

%FIGURE 18

\clearpage
%FIGURE 19

%FIGURE 20

%FIGURE 21

%FIGURE 22

%Cumulative energy
\clearpage
%FIGURE 23

%FIGURE 24

%FIGURE 25

%FIGURE 26

\clearpage
%FIGURE 27

%FIGURE 28

%FIGURE 29

%FIGURE 30

\clearpage
%FIGURE 31

%FIGURE 32

%FIGURE 33

%FIGURE 34

\clearpage
%FIGURE 35

%FIGURE 36

%FIGURE 37

%FIGURE 38

\clearpage
%FIGURE 39

%FIGURE 40

%FIGURE 41

%FIGURE 42

%%%%%%%%%%%%% Post-diction %%%%%%%%%%%%%%%%%%%%%%%%%%%
%FIGURE 43

%FIGURE 44

%FIGURE 45

%FIGURE 46

%FIGURE 47

\end{document}